\documentstyle[preprint,floats,aps,epsf,epsfig,prb]{revtex}

\newlength{\bxwidth}\bxwidth=2.5 truein

\newcommand\frm[1]{\epsfig{file=#1,width=\bxwidth}}
\newlength{\fight}
\newcommand\ltdash{\raise-1.8pt\hbox{$\scriptscriptstyle |$}}
\newcommand \beq  {\begin{equation}}
\newcommand \eeq  {\end{equation}}
\newcommand \bea {\begin{eqnarray} }
\newcommand \eea {\end{eqnarray}}

\newcommand \rarrow{\rightarrow}
\newcommand \dg{^{\dagger}}
\newcommand \si { \sigma}
\newcommand \ra { \rangle}
\newcommand\la{\langle}

\begin{document}
\draft
\title{
Strong Coupling Approach to the Supersymmetric Kondo Model.
}
\author{  P.
Coleman
$^{1}$ , C. P{\'e}pin $^2$ and A. M. Tsvelik
$^{2}$}
\address{$1$ Materials Theory Group,
Department of Physics and Astronomy,
Rutgers University,
136 Frelinghausen Road,
Piscataway, NJ 08854,
USA
}
\address{$2$ Department of Physics, 
Oxford University, 
1 Keble Road, 
Oxford OX1 3NP,UK }
\maketitle
\date{\today}
\maketitle
\begin{abstract}
We carry out the strong coupling expansion for  the $SU(N)$ Kondo model where
the impurity spin is represented by a L-shaped Young tableau. Using
second order perturbation theory around the strong coupling fixed
point it is shown that when the antisymmetric component of the
Young-tableau contains more than $N/2$ entries, the strong-coupling
fixed point becomes unstable to a two-stage Kondo effect. 
By comparing the strong coupling results obtained here
with the result using  a supersymmetric large  $N$ expansion, we
are also able to confirm the validity of the 
the supersymmetric formalism for mixed symmetry Kondo models.
\end{abstract}
\vskip 0.2 truein
\newpage

In this paper we present a strong-coupling treatment of a single-impurity
Kondo model where the spin is a higher representation of the group SU(N).
We consider spin representations
that can be tuned continuously from being antisymmetric to being fully symmetric. 
This work is motivated in part by a desire to understand how the presence
of strong Hund's interactions between electrons modify the spin quenching
process. These issues become particularly important in heavy electron
systems, where the localized electrons can be subject to Hund's
interactions which far exceed their kinetic energy. \cite{cox94,norman}

The  basic model of interest is the single-site Kondo  model,
described by the Hamiltonian
\bea
H&=& \sum \epsilon_{k} c^{\dagger}_{k \si}
c_{k \si} + 
\frac{J}{N}c\dg_{\alpha}(0)\pmb{$\Gamma$}_{\alpha \beta}c_{\beta}(0) \cdot
{\bf S}.
\eea
Here, the spin sums run over $N>1$ possible values, 
$c\dg_{\alpha}(0)= \frac{1}{\sqrt{n_s}}\sum_k c\dg_{k\alpha}$ creates
an electron at the origin, $n_s$ is the number of sites.
The matrices $\pmb{$\Gamma$}= ( \Gamma^1, \Gamma^2 \dots)$ are the 
form a basis of $N^2-1$ traceless SU(N) matrices, where ${\rm
  Tr}[\Gamma^a\Gamma^b] = \delta^{ab}$ and ${\bf S}=(S^1,S^2 \dots )$ is a spin describing
a particular representation of SU(N).   
The above model was first
derived by Coqblin and Schrieffer~\cite{coqschrieff}, who showed that a rare earth
ion containing a single, spin-orbit coupled  f-electron corresponds
to the above model, with $N=2j+1$, and the spin in the fundamental
representation of SU(N).  

When we come to consider more complex local moment systems, we need
to consider the atomic spins formed by combining more than one elementary
spin.  Previous treatments of
this model have considered local moments described by 
symmetric or  antisymmetric representations
of SU(N), denoted
by the Young tableaux~\cite{yt}
\setlength {\unitlength}{0.004\textwidth}
\begin{center}
\begin{picture}(200,90)(0,0)
\put(0,80){(a) Antisymmetric}
\put(20,60){\vector(0,1){10}}
\put(19,50){$n_f$}
\put(20,40){\vector(0,-1){10}}
\put(45,30){\framebox(10,40){}}
\put(45,40){\line(1,0){10}}
\put(45,60){\line(1,0){10}}
\put(50,45){.}
\put(50,50){.}
\put(50,55){.}
\put(125,80){(b) Symmetric}
\put(120,50){\framebox(60,10){}}
\put(130,50){\line(0,1){10}}
\put(170,50){\line(0,1){10}}
\put(140,55){.}
\put(150,55){.}
\put(160,55){.}
\put(140,40){$n_b\equiv 2S$}
\put(138,42){\vector(-1,0){18}} 
\put(163,42){\vector(1,0){17}} 
\end{picture} 
\end{center}
The first representation describes $n_f$ elementary spins that
have been combined into a purely antisymmetric spin wavefunction;
the second describes $n_b\equiv 2S $ elementary spins that have been
combined into a purely symmetric spin wavefunction.  These two
representations are of particular interest because the former
can be described by a combination of $n_f$ spin, or ``Abrikosov'' pseudo 
fermions, whereas
the latter can be described by a combination of $n_b\equiv 2S$ ``Schwinger
bosons", and is the natural generalization of  spin-S to $SU(N)$. 
For a given number of spins, 
the antisymmetric and symmetric spin combinations represent two extremes
where the ``Casimir'' ${\bf S}^2$ attains its extremal values. 
Loosely speaking, the antisymmetric and  symmetric
spin representations are the combination of spins with the 
smallest and largest total spin,
respectively.  The symmetric spin configuration can thus be thought
of as a state where a large Hund's coupling has maximized the total spin.

The ``column -shaped'' representations of the SU (N) Kondo model have been
considered by previous authors in various contexts.   The SU (N) Kondo model
in its fundamental representation has been treated using  both Bethe
ansatz\cite{andreiA,tsvelikW}
and 
later in the context of a path integral large 
$N$ expansion\cite{read}.  
Later work lead to the realization that higher antisymmetric 
``column'' representations of the same
model are required for a well-defined large N expansion
\cite{andreicoleman,coleman}. The introduction of conformal field
theory \cite{affleck} showed how this could be
used to compute spin correlation functions in both the single and
multi-channel models. More recently, 
row\cite{parc1} and then column\cite{parc2}
representations of SU (N) of the multi-channel Kondo model have also
been considered using both large $N$ and conformal field theory
methods.  In all of these cases, screening of the local moment
involves a single Kondo energy scale, independently of whether the moment
is screened, underscreened or overscreened. 

In this paper, we are interested in a new class of ``L-shaped'' spin
representations composed of spins which interpolate between row and
column
representations, as shown in Fig 1.  below
\begin{figure}[here]
\epsfysize=0. \textwidth 
\centerline{\epsfbox{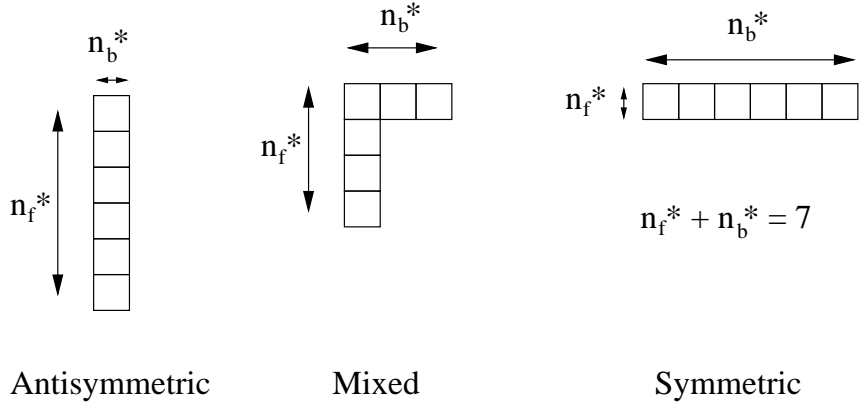}}
\vskip 0.3truein
\protect\caption{ A sequence of L-shaped Young tableaux  which interpolate
betwen and antisymmetric and symmetric representation.
Each tableau has six boxes, corresponding to six elementary spins. }
\label{lshaped}
\vskip 0.1truein
\end{figure}
This family of representations enables us to examine
the effect of progressively turning on the Hund's interaction
the effect of progressively increasing the strength of the Hund's
interaction between the constituent spins inside an atom. 
In are a real multi-electron local moment, such as a $U^{3+}$ ion,
containing three localized $f-$ electrons, the Hund's interaction
also imposes the crystalline symmetry, leading to a model with a
far lower symmetry. Our toy representation enables us to separate 
the leading order effect of the Hund's interaction
from the additional complications of lowering the symmetry. 

In a previous paper~\cite{firstpaper}, we showed
that the above mixed symmetry representations of SU(N) spins can be desribed
using a ``super-symmetric spin'' representation. In particular,
if $b\dg_{\alpha}$ and $f\dg _{\alpha}$  ($ \alpha = (1,N)$)
 are bose and Fermi creation
operators, respectively, then a mixed symmetry representation of
is obtained by writing the spin ${\bf S}$ as a sum of a bosonic
and a fermionic spin
\bea
{\bf S} = b\dg_{\alpha} \pmb{$\Gamma$}_{\alpha \beta} b_{\beta}
+f\dg_{\alpha} \pmb{$\Gamma$}_{\alpha \beta} f_{\beta}
\eea
where $\pmb{$\Gamma$}\equiv ( \Gamma^1, \dots
\Gamma^{M})$ represents the $M= (N^2-1)/2$ independent
SU(N) generators. In this way, the spin representation
combines aspects of the bosonic ``Schwinger boson'' representation
of spins and the fermionic ``Abrikosov pseudo-fermion'' representation
of spins. 
This spin operator commutes with the
the operators
\bea
{\theta }\dg =\sum_{\beta = 1, N}
f\dg
_{\beta}b_{\beta}, \qquad { \theta}  = \sum_{\beta = 1, N} b\dg_{\beta}f_{\beta}.
\eea
These operators interconvert the bose and fermion fields and form 
the generators of the supergroup $SU(1|1)$.  An
irreducible L-shaped representation of $SU(N)$ is obtained by imposing
two constraints
\bea
\hat n_f + \hat n_b &=& Q,\cr
\hat n_f - \hat n_b + \frac{1}{Q}[\theta\dg, \theta] &=& Y,
\eea
where $Q$ and $Y$ are determined from the Young tableau via
the relations $Q= n_f^* + n_b^*$ and $Y= n_f^* - n_b^*$. 
These constraints also commute with the generators $\theta$ and $\theta\dg$,
so the entire spin representation is supersymmetric. 

This mixed representation of the spin-operators allows a consideration
of the properties of Kondo models with L-shaped Young-tableaux 
by developing a supersymmetric field theory and then carrying out
a large-N expansion. 
In the conventional one-channel Kondo model, the spin is screened
from $S$ to $S-\frac{1}{2}$ by a process that is characterized by a single
temperature scale, the Kondo temperature. In the classic picture, this
leads to a screening cloud of dimension $l=v_F/ T_K$, where $v_F$ is the
Fermi velocity of electrons in the conduction sea and $T_K$ is the single
Kondo temperature.
One of the unexpected results of the new analysis,
was that under some circumstances
a local moment under the influence of a Hund's interaction
can undergo a two-stage
Kondo effect associated with two separate Kondo temperatures
$T_{K1}$ and $T_{K2}$, leading to a screening cloud with ``internal structure'',
characterized by two screening length scales 
\bea
l_1= \frac{v_F}{T_{K1}}, 
\qquad l_2= 
\frac{v_F}{T_{K2}}
\eea 
where $v_F$ is the Fermi velocity of the electrons,
as illustrated in Fig. 2. 
\begin{figure}
\epsfysize=0.3 \textwidth 
\centerline{\epsfbox{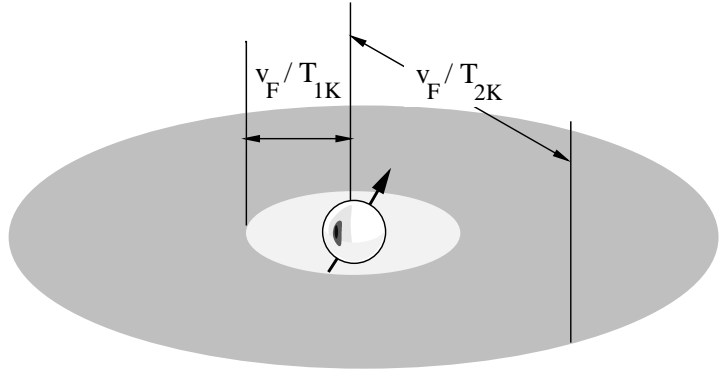}}
\vskip 0.3truein
\protect\caption{ Illustrating a local moment that is screened
by a two-stage screening process in a single scattering channel.
The screening cloud contains a ``shell structure'', with an inner
and outer conduction electron cloud which reduce the total moment
from $S$ to $S^*=S-1$.}
\label{quench}
\vskip 0.4truein
\end{figure}
In the language of the renormalization
group, a new fixed point occurs where the impurity is screened in two stages
reducing the effective spin from $S$ to 
to $S^*=S-1$, where $2S=n_{b}^{*}$ and  $2S^{*}$ are the
width of the 
Young Tableau for the bare and the partially screened local
moment, respectively.
In this paper
we provide a complimentary ``strong-coupling'' treatment of the same
model. Our results 
confirm the key results obtained in the large $N$ expansion, providing
an important check on the validity of the supersymmetric field theory. 

It is well known that under the renormalization group, 
the antiferromagnetic Kondo model renormalizes
towards  strong coupling~\cite{anderson,nozieres}.  The weak-coupling
beta function for the Kondo model is independent of representation,
and for a single channel model,  takes the form
\bea
\beta(g)=\frac{d g(\Lambda)}{d\Lambda}= - g^2 + \frac{g^3}{N} +O (g^{4})
\eea
where $g= J \rho$ and $\rho$ is the conduction electron density of states.
For the simplest one-channel Kondo models, the coupling constant $g$ flows
smoothly from the unstable weak-coupling fixed point, to 
a stable strong coupling fixed point. 
However, in 
certain cases
the strong coupling fixed point is itself unstable. 
The most famous example of such
behaviour is the two-channel Kondo model, where the flow to
strong coupling is intercepted by an intermediate coupling fixed
point that is characterized by a non fermi liquid properties~\cite{blandin,tsvelik,andrei}.
The two stage Kondo effect discussed here
is another example of an instability at strong coupling. As we shall see,
the strong coupling fixed point becomes unstable to a second Kondo effect
at an exponentially smaller temperature.

The classic analysis~\cite{blandin} of the stability of the
strong coupling fixed point of a Kondo model follows Wilson's method~\cite{wilson}
of formulating the Kondo model on a lattice:
 \bea
H=  -t \sum_{n\geq 1, \alpha} \left [  c\dg_{ \alpha} (n+1)c_{ \alpha
}(n) +{\rm H.c}\right]+
 \frac{J}{N} c\dg_{\alpha} (0) \pmb{$\Gamma$}_{\alpha
\beta}c_{\beta} (0)
\cdot
{\bf S} 
\eea
The strong-coupling fixed point is obtained by first setting $t=0$, and
solving for the ground-state of the one-site problem
\bea
H_K= \frac{J}{N}c\dg_{\alpha}(0)\pmb{$\Gamma$}_{\alpha \beta}c_{\beta}(0) \cdot
{\bf S}
\eea
which leads to a partially screened local moment with spin $S^*$. 
When a finite $t/J$ is restored, virtual charge fluctuations
of electrons onto, and off site 0 induce an effective interaction between
the spin density at site $1$ and the residual moment, given by
\bea
H^{(1)}= J^* {\bf S}^*\cdot c\dg_{\alpha} (1) \pmb{$\Gamma$}_{\alpha
\beta}c_{\beta} (1)
\eea
where $J^*= O(t^2/J)$ determines the strength of the 
coupling between the residual local moment and the conduction electron at
site $1$. 
The stability of the strong
coupling fixed point is determined by the sign of $J^*$. 
If this coupling is ferromagnetic ($J^* <0$), then a residual ferromagnetic
Kondo effect with the low energy electrons causes $J^*$ to scale
logarithmically to zero, decoupling the residual spin from the conduction
sea and stabilizing the strong-coupling fixed point.
Conversely if the effective coupling is
antiferramagnetic ($J^*>0$), then the effective model flows to strong
coupling and the strong coupling fixed point of the initial Kondo
model becomes unstable.

To get a better idea of how antiferromagnetic coupling can arise in our
{\sl single-channel} model, 
consider a local moment, described by an L-shaped
representation of SU(N), denoted by  the Young Tableau
\setlength {\unitlength}{0.005\textwidth}
\bea
{\bf S}=\parbox{50\unitlength}{
\begin{picture}(50,40)(0,0)
\multiput(20,25)(5,0){5}{\framebox(5,5){}}
\put(28,37){\vector(-1,0){8}}
\put(37,37){\vector(1,0){8}}
\put(30,35){$2S$}
\multiput(20,20)(0,-5){4}{\framebox(5,5){}}
\put(13,22){\vector(0,1){8}}
\put(10,17){$n_f^*$}
\put(13,13){\vector(0,-1){8}}
\end{picture} }
\eea
where we have replaced 
$n_{b}^{*}=2S$.
In the 
ground-state of the strong-coupling Hamiltonian
$
H_K
$,
electrons form a singlet with the fermionic part of the spin creating a partially screened
moment,  denoted by a Young-tableau with a completely filled first row.
\bea
{\bf S}^*=(\pmb{$\Gamma$}_{e}(0)+{\bf S})=\parbox{50\unitlength}{
\begin{picture}(50,60)(0,0)
\multiput(20,45)(5,0){5}{\framebox(5,5){}}
\put(28,57){\vector(-1,0){8}}
\put(37,57){\vector(1,0){8}}
\put(30,55){$2S$}
\multiput(20,40)(0,-5){7}{\framebox(5,5){}}
\put(13,35){\vector(0,1){15}}
\put(10,30){$N$}
\put(13,26){\vector(0,-1){15}}
\put(20,12){$c_0$}
\put(20,17){$c_0$}
\put(20,22){$c_0$}
\end{picture} }
\equiv\parbox{50\unitlength}{
\begin{picture}(50,15)(0,0)
\multiput(20,5)(5,0){4}{\framebox(5,5){}}
\put(27,17){\vector(-1,0){7}}
\put(26,17){\vector(1,0){14}}
\put(22,23){$2S-1$}
\end{picture}
}
\eea
where in this example we have taken $N=8$.
Since the first column of the tableau is a singlet (with $N$ boxes), it 
can be removed from the tableau,leaving behind a partially screened
spin $S-1/2$, described by a row with $2S-1$ boxes.
If we now couple the electron at the origin with electrons at site `1'
via a small hopping matrix element $t<<J$, then the virtual charge fluctuations
of electrons in and out of the singlet at the origin will lead to a residual
coupling
between the partially screened moment and the electrons
at the neighboring site '1'
\bea
H^{(1)}= J^* {\bf S}^*\cdot c\dg_{\alpha} (1) \pmb{$\Gamma$}_{\alpha
\beta}c_{\beta} (1)
\eea
where $J^*\sim t^2/J$. 
In the  SU(2) Kondo model, only electrons  parallel
to the residual moment ${\bf S}^*$ can hop onto the origin, which gives
rise to a ferromagnetic coupling $J^*<0$.  In the 
SU(N) case, electrons can hop  provided they are not in the same
spin state as electrons  at the origin.
The sign of the coupling $J^*$  depends on the number of conduction
electrons $n_c= N-n_f^*$, bound at the origin. For example, if $n_f^*=1$,
so that $n_c=N-1$ in the ground state, 
electrons hopping onto the origin will have to be parallel to the residual
spin, so in this case the coupling {\sl is} ferromagnetic, $J^*<0$. 
By contrast,
if $n_c<<N$, there are many ways
for the electron to hop onto the origin with a spin component that is
different to the residual moment, so the residual
interaction will be antiferromagnetic, $J^*>0$.   
In this case, the 
the strong-coupling fixed point becomes unstable, and a second-stage Kondo effect
occurs, binding a further 
$N-1$ electrons at site "1" to form a state represented by the tableau
\begin{equation}
{\bf S}^{**}=(\pmb{$\Gamma$}_{e_0}+\pmb{$\Gamma$}_{e_1}+
{\bf S})=\parbox{50\unitlength}{
\begin{picture}(50,60)(0,0)
\multiput(20,45)(5,0){5}{\framebox(5,5){}}
\put(28,57){\vector(-1,0){8}}
\put(37,57){\vector(1,0){8}}
\put(30,55){$2S$}
\multiput(20,40)(0,-5){7}{\framebox(5,5){}}
\put(13,35){\vector(0,1){15}}
\put(10,30){$N$}
\put(13,26){\vector(0,-1){15}}
\put(21,12){$c_0$}
\put(21,17){$c_0$}
\put(21,22){$c_0$}
\multiput(25,40)(0,-5){7}{\framebox(5,5){}}
\put(26,12){$c_1$}
\put(26,17){$c_1$}
\put(26,22){$c_1$}
\put(26,27){$c_1$}
\put(26,32){$c_1$}
\put(26,37){$c_1$}
\put(26,42){$c_1$}
\end{picture}}
\equiv
\parbox{50\unitlength}{
\begin{picture}(50,15)(0,0)
\multiput(20,5)(5,0){3}{\framebox(5,5){}}
\put(28,15){\vector(-1,0){8}}
\put(25,15){\vector(1,0){10}}
\put(20,20){$2S-2$}
\end{picture}
}
\end{equation}
which corresponds to a residual spin $S^{**}=S-1$. 
This final configuration is stable, because  an electron at site
"2" can only hop onto site "1"  if it is parallel to the
unquenched moment, so the residual interaction between site ``2'' and site ``1'' will be ferromagnetic. 

To examine the stability of the strong coupling fixed point in detail,thisail, this paper
follows the method of Nozi{\`e}res and Blandin~\cite{blandin},  using 
second order perturbation
theory about the strong-coupling fixed point to  determine
the sign of the residual interaction between the
unscreened soin and the bulk of conduction electrons. If
$J^* <0$ the residual coupling is ferromagnetic and the strong
coupling fixed  point is stable: the impurity residual spin is
$S^*=S-1/2$ after screening. If $J^*>0$ the residual interaction is
antiferromagnetic and the strong coupling fixed point becomes
unstable, leading to the two-stage Kondo effect $S^*=S-1$.
In parallel with this approach, we carry out a large $N$
treatment of the strong-coupling limit,
using the technique developped in our previous
paper. By comparing the two techniques we are able to confirm
the validity of the field theoretic approach developed in our earlier
work. 
Both methods are able to confirm that for $N>2$, 
$J^*$ changes sign when the number of bound-conduction electrons is less than
$N/2$, and in the large $N$ limit is given by
\bea
J^*= - \frac{t^2}{J (1 -\tilde n_f)\tilde n_f}
\left[\frac{\frac{1}{2} - \tilde n_f}{(1 -
\tilde n_f + \tilde n_b)}\right]
\eea
where $\tilde n_f= n_f/N$ and $\tilde n_b = 2S/ N$.

The table below gives a shows the main results, comparing 
the strong coupling and large $N$ expressions for the ground-state energy
$E_g$, the excitation energie $\Delta E^{\uparrow \uparrow }$ and
$\Delta E^{\uparrow\downarrow}$
to add an electron
to the ground state in a spin configuration that is
``parallel''  or ``anti-parallel'' to the residual spin. 
The last row compares the effective coupling constant $J^*$ 
calculated by  both methods. In these expressions, we denote
$\tilde n_f = n_f^*/N$, $\tilde n_b= n_b^*/N$, $q= Q/N$.

\begin{tabular}{||c||c|c||} \hline \hline
& Strong Coupling & Large N \\ \hline \hline
$\begin{array}{c}  \mbox {Ground State Energy} \\
E_g \end{array}$ & $  -NJ \left [ ( 1-\tilde n_f ) (\tilde n_f
+ q /N  \right ] $ & $ - N J (1-\tilde n_f){\tilde n}_f $ \\ \hline
$ \begin{array}{c} \Delta E^{\uparrow\downarrow}= E^{\uparrow\downarrow}(n_e+1) - E_g \\
\quad \end{array}$ & $ J ( 1-\tilde n_f -q/N)$ 
& $J (1- {\tilde n}_f )$ \\ \hline
$ \begin{array}{c}\Delta E^{\uparrow\uparrow}= E^{\uparrow\uparrow}(n_e+1) - E_g
\\
\qquad \end{array}$ & $J (1- \tilde n_f+ \tilde n_b -q/N) $ & $ J(1- {\tilde n}_f +
{\tilde n}_b )$ \\ \hline
$\begin{array}{c}
\mbox{Effective coupling}\\
\mbox{constant}\end{array}$ & $  \begin{array}{l}  \frac{t^2}{(
2 S ) J } \left [ \frac{n_f N}{ (N-1)(  N - n_f^* -Q/N)} \right .\\
  - \frac{Q N}{(Q+N n_f^*) ( 1 + N + Q - 2 n_f^* - Q/N
)} \\  \left .
 - \frac{ ( N-n_f^* )(Q-n_f^*) N}{ (N-1) ( Q+N -n_f^*) (n_f^* +Q/N)}
\right ] 
\end{array}  $  & $ {\displaystyle - \frac{t^2 (1/2 - {\tilde n}_f
)}{J ( 1-{\tilde n}_f) {\tilde n}_f ( 1- {\tilde n}_f + {\tilde n}_b
)}} $ \\ \hline \hline
\end{tabular}

\section{The one site impurity problem}

To begin, we start with the one-dimensional rendition
of the SU(N) Kondo model, 
 \bea
H=  -t \sum_{n\geq 0, \alpha} \left [  c\dg_{ \alpha} (n+1)c_{ \alpha
}(n) +{\rm H.c}\right]+ 
 \frac{J}{N} c\dg_{\alpha} (0) \pmb{$\Gamma$}_{\alpha
\beta}c_{\beta} (0)
\cdot
{\bf S} 
\eea
where
$c\dg_{\alpha} (j) $ creates an electron at the site of the local moment, ${\bf S}$ is
the spin of the local moment and $\pmb{$\Gamma$}_{\alpha \beta}$ is a
generator of the SU(N) group. For simplicity we have put the impurity
at the begining of the chain. The spin of
the impurity is represented by an L-shaped Young tableau while the
conduction electron is represented with a single box.

\begin{figure}
\epsfysize=0.3 \textwidth 
\centerline{\epsfbox{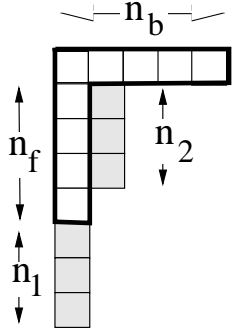}}
\vskip 0.3truein
\protect\caption{Screening of the impurity. $n_1$ and $n_2$ denote the
number of conduction electrons screening the impurity in the first and
second column of the Young tableau.}
\label{singlesite}
\end{figure}

In the strong coupling limit, only the local part of the hamiltonian
is relevant and the problem becomes single sited. In order to
determine what is the ground state of the single site problem, we
consider the impurity described by a Young Tableau with $Q$ boxes.
In order to simplify further calculations, we denote by $n_b$ the
number of boxes in the row and by $n_f$ the number of boxes in the
column minus one. ( This will correspond in the next section to put
some bosons in the row as well as in the corner and to fill the
remaining of the column with fermions.)
In the strong coupling limit, the conduction electrons will be trapped
at the impurity site, screening the impurity. We note $n_1$ and $n_2$
the number of conduction electrons screening the impurity respectively
in the first and second column of the Young Tableau (see
Fig.~\ref{singlesite}).

The energy of the ground state can be expressed in terms of second
order Casimirs

\bea
E & = & \frac{J}{N} {\bf S}_c \cdot {\bf S}_i \nonumber \\
  & = & \frac{J}{2 N} \left [ {\bf S}^2_{tot} - {\bf S}^2_{i} -{ \bf
  S}^2_{c} \right ] \ ,
\eea
where ${\bf S}^2_{tot}$ is the second order Casimir of the impurity
screened by the $n_1 + n_2$ conduction electrons; ${\bf S}^2_{i}$ is
the Casimir of the free impurity and ${ \bf
  S}^2_{c}$ is the Casimir of the conduction electrons which screen
the impurity.

If we normalize the generators of the fundamental representation of
SU(N) according to $Tr [ \Gamma^{\alpha} \Gamma^{\beta} ] =
\delta^{\alpha \beta} $, then the expression for the 
Casimir of an arbitrary irreducible representation is~\cite{zarand}
\bea
{\bf S}^2 = \frac{Q ( N^2-Q )} {N} + \sum_{j=1,N} m_j \left ( m_j+1-2
j \right ) \ ,
\eea where $m-j$ is the number of boxes in the $j$-th row from the
top.
The energy of the ground state is then given by
\bea
E & = & \frac{J}{2 N} \left [ 2 n_2 \left ( N- n_2 \right ) + \left (
n_f  + n_1 - n_2 \right ) \left ( N - n_f - n_1 - n_2 -1 \right )
 \right . \nonumber \\
  & - &  \left . \frac{1}{N} \left ( Q + n_1 + n_2 \right )^2 - n_f \left ( N - n_f -1 \right ) + \frac{1}{N} Q^2
\right . \nonumber \\
  & - & \left . \left ( n_1 + n_2 \right ) \left ( N +1 - n_1 - n_2 \right ) +
\frac{1}{N} \left ( n_1 + n_2 \right )^2 \right ] \ .
\eea

\begin{figure}[here]
\epsfysize=0.4\textwidth
\centerline{\epsfbox{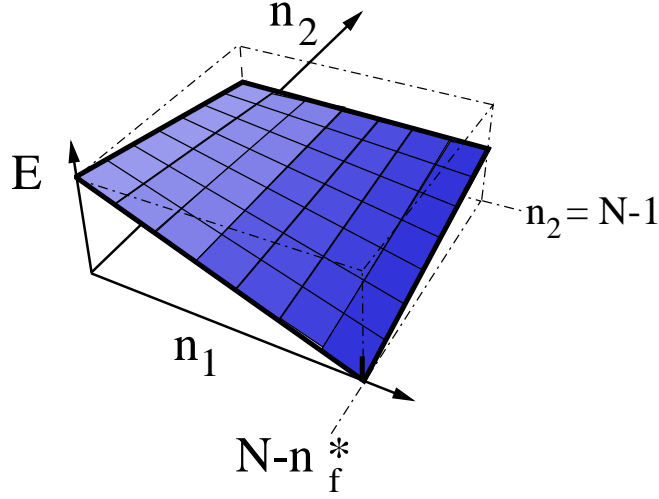}}
\vskip 0.1truein
\protect\caption{Single impurity energy as a function of the
parameters $n_1$ and $n_2$. We see that the minimum is given when
$n_1$ is maximum ($n_1 = N-n_f -1$) and $n_2$ is minimum ( $n_2=0$).}
\label{tortellini}
\end{figure}

The energy
in the $[n_1,n_2]$ plane has the form represented in
figure~\ref{tortellini}. 
In the
range of parameters we are interested in ($n_1 \in [0, N - n_f - 1]$,
$n_2 \in [0, N-1]$),  only one minimum remains and the ground state of the problem is found to be 

\beq
\begin{array}{cc}
n_1 = N - n_f -1 \ \ \ ; &  n_2 = 0 
\end{array}
\eeq
which corresponds to the usual one-stage Kondo model where the
impurity spin is screened by $1/2$.

\begin{figure}
\epsfysize=1.5in 
\centerline{\epsfbox{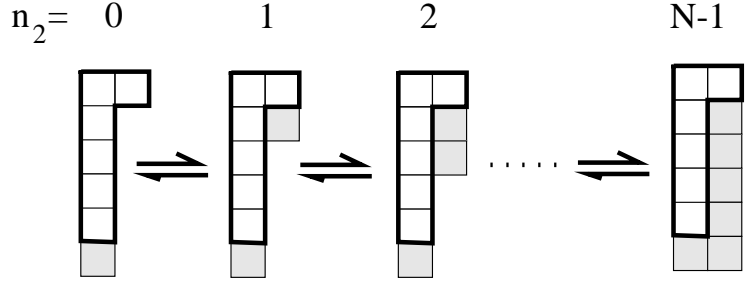}}
\vskip 0.1truein
\protect\caption{In the special case where $Q/N=$k is an integer
and $n_1=$k, it costs no energy to add additional electrons to the
ground-state and the fixed point behavior of this particular
Kondo model will then involve valence fluctuations between the
different degenerate configurations. The figure illustrates
the situation where $Q=N=6$ and 
$n_1=1$.
}
\label{degenerate}
\end{figure}

A special exception to this case occurs when  $Q/N=k$ is an integer,
when, if $n_{1}=Q/N=k$, ${
\frac{\partial E}{\partial n_2} =0}$ { for all} $n_2$.
The point $n_1= Q/N=k, n_{2}\in [1,N-1]$ corresponds to a line of degenerate 
ground states. An example is illustrated on
figure~\ref{degenerate}. 
In this case, the strong coupling fixed point is also unstable, but
the fixed point physics will be governed by valence fluctuations
between
the degenerate states of different $n_{2}$.  We shall exclude this
special case from the discussion here, leaving it for future work.

\section{Second order perturbation theory around the strong coupling
fixed point}

In the leading order in $\frac{1}{J}$ two processes of excitation
appear. Suppose the screened impurity is at site ``zero''. Then one
electron from site ``one'' can hop briefly to site zero, then return  (process
1).
\begin{figure}[here]
\epsfysize=1.5in 
\centerline{\epsfbox{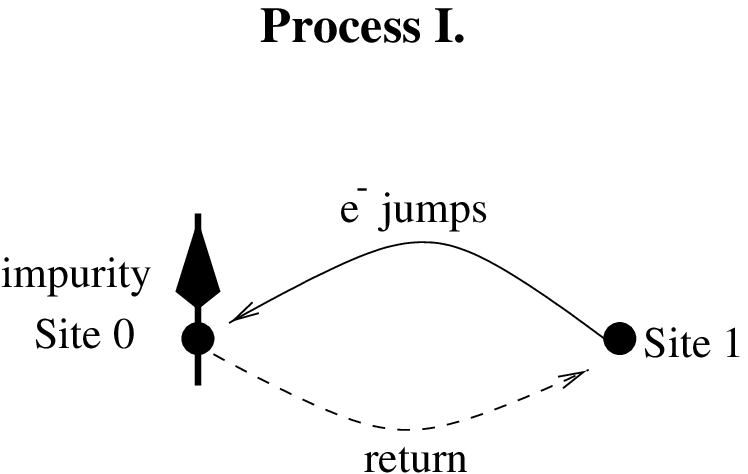}}
\vskip 0.1truein
\protect\caption{Process I: an  electron  
makes a virtual incursion from site 1 onto site 0. }
\label{process1}
\end{figure}
Alternatively one electron from site ``zero'' can make a virtual
hop to site one and return (process 2).
\begin{figure}[here]
\epsfysize=1.5in 
\centerline{\epsfbox{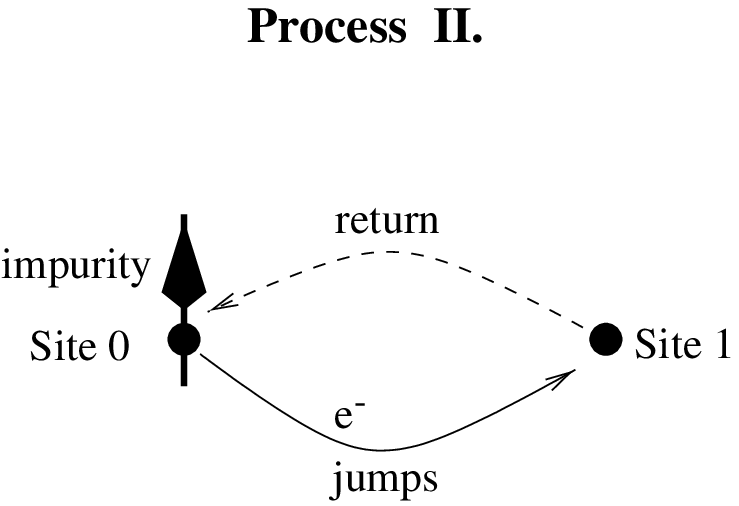}}
\vskip 0.1truein
\protect\caption{ Process II : an electron at site zero 
makes a virtual incursion to 
to site 1.}
\label{process2}
\end{figure}
In the first
process the intermediate state has one more electron at site one ($
n_c +1$), where $n_c$ is the number of conduction electrons which
screen the impurity in the ground state). We note this state $ |GS+1,
0 \rangle $. In the second process of excitations the intermediate
state has $n_c - 1 $ electron at site zero and two electrons at site
one. We note it $ |GS -1, 2 \rangle $. If we call $ | GS, 1
\rangle $ the initial state ( the site 0 has the impurity in the
ground state and the site 1 has one electron ),then using second order
perturbation theory, the energy shift due to the perturbation is given
by
\bea
\Delta E & = & t^2 \frac{ | \langle GS + 1 , 0 | c^\dagger_1 c_2 | GS,1
\rangle | ^2 }{ E(n_{e}) - E(n_{e}+1) }  \nonumber \\
   & + & t^2 \frac{ | \langle GS -1, 2 | c^\dagger_2 c_1 | GS,1 \rangle
   |^2 }{ E(n_{e}) -E(n_{e}-1) } \ ,
\eea where $E(n_{e})$ is the energy of the initial state; $E(n_{e}+1)$ is the
energy of the intermediate state in process 1 and $E(n_{e}-1)$ is the
energy of the intermediate state in process 2.

\begin{figure}[here]
\epsfxsize=0.8\textwidth
\centerline{\epsfbox{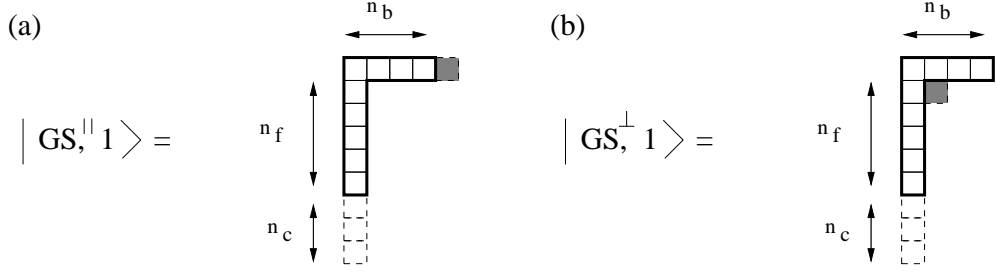}}
\vskip 0.1truein
\protect\caption{{\bf (a)}
``Parallel'' configuration: the spin of the electron at site 1 is 
symmetrized with the impurity spin.
{\bf (b)}
``Anti-Parallel'' configuration: the spin of the electron at site 1 is 
anti-symmetrized with the impurity spin.
}
\label{both}
\end{figure}

Now for each process of excitation the spin of the electron at site 1 can
is either  symmetrically(Fig. ~\ref{both} (a)), or antisymmetrically
(Fig. ~\ref{both} (b)) correlated with the spin
at the impurity.  For SU(2) spins, this corresponds to a spin that is
either ``parallel'' or ``anti-parallel'' to the impurity spin, and 
we shall adopt the same convention for the SU(N) case. 
There are 
two corresponding possibilities for the energy shifts
\beq
 \left \{ \begin{array}{l}
\Delta E^{\uparrow\uparrow} = {\displaystyle \frac{ M^{\uparrow\uparrow}_{(1)} }{
E(n_{e})-E^{\uparrow\uparrow}(n_{e}+1)} + \frac{ M_{(2)}^{\uparrow\uparrow} }{ E(n_{e})-E(n_{e}-1)}} \\
\Delta E^{\uparrow \downarrow} = {\displaystyle \frac{ M^{\uparrow \downarrow}_{(1)} }{
E(n_{e})-E^{\uparrow \downarrow}(n_{e}+1)} + \frac{ M_{(2)}^{\uparrow \downarrow} }{ E(n_{e})-E(n_{e}-1)}}
\end{array} \right . \ ,
\eeq 
\noindent where for example $ M_{(1)}^{\uparrow\uparrow} $ and
$ M_{(1)}^{\uparrow\downarrow} $ 
are the matrix elements for 
process one in the parallel 
and antiparallel configurations, respectively.

Supposing now that the effective Hamiltonian around the strong
coupling fixed point takes the form
\bea
H^{(1)}= J^* {\bf S}^*\cdot c\dg_{\alpha} (1) \pmb{$\Gamma$}_{\alpha
\beta}c_{\beta} (1)
\eea
by computing the energy difference between the parallel and
anti-parallel configurations, the effective Kondo
coupling constant is then 
\beq\label{eq2s}
\Delta E^{\uparrow\uparrow} - \Delta E^{\uparrow \downarrow} = J_{eff} ( 2S) 
.
\eeq
Thus by evaluating
the energy shifts in the second order perturbation theory we are able to determine the sign and hence the stability of the fixed point. 

\subsection{Evaluation of the energies }

First, consider the energy of the ground state at site one. It is given
by
\beq
\label{meanfield}
E(n_{e}) = - \frac{J}{N} \left [ \left ( N -1-n_f \right ) \left ( n_f +
(N+Q)/N \right ) \right ] \ .
\eeq
Suppose now one electron jumps from site 1 to site 0 in the ``parallel''
state, as in 
figure~\ref{both} (a), 
the 
the energy of the intermediate state is 
\[
E^{\uparrow\uparrow}(n_{e}) = \frac{J}{2 N} \left[ {\bf S}^2_{N+1} - {\bf
S}^2_{imp} - {\bf S}^2_{el} \right ] \] where ${\bf S}^2_{N+1}$ is the
Casimir of the intermediate state; ${\bf S}^2_{imp}$ is the Casimir of
the imurity before screening and ${\bf S}^2$ is the Casimir of all
($n_c+1$) conduction electrons involved into the intermediate
state. The resulting energy is given by
\[
E^{\uparrow\uparrow}(n_{e}+1) = - \frac{J}{N^2} \left [ n_f \left ( N^2 -n_b -n_f - N
n_f \right ) \right ] \ .
\]
Similarly if the starting spin configuration is the ``anti-parallel''
one, as in  \ref{both} (b),  the energy of the
intermediate state is 
\[
E^{\uparrow \downarrow}(n_{e}+1) = - \frac{J}{N} \left [ n_b \left ( 1-n_f/N \right ) + n_f \left
(N- n_f - n_f/N \right ) \right ] \ .
\]

For process 2, the energy of the intermediate state does not depend on
whether the spin configuration is parallel or anti-parallel, and is given
by
\[
E(n_{e}-1)- E(n_{e}) =  \frac{J}{N} \left ( n_f + 1 + Q /N \right ) \ . \]
In conclusion the two energy differences associated with process 1 are
given by 
\beq
\label{shifts}
\left \{ \begin{array}{l}
 E^{\uparrow \downarrow}(n_{e}+1) - E(n_{e}) = \frac{J}{N} \left ( N - n_f^* - Q/N \right ) \\
 E^{\uparrow\uparrow}(n_{e}+1) -E(n_{e}) = \frac{J}{N} \left ( N -n_f^* +n_b^*
- Q/N \right )
\end{array} \right . 
\eeq 
where we have replaced $n_f+1 \rarrow n_f^*$ and $n_b\rarrow n_b^*$. 
The excitation energy associated with process 2 is
\beq
E(n_{e}-1)- E(n_{e}) =  \frac{J}{N} \left ( n_f^* + Q /N \right )
\eeq
for both spin configurations.

\subsection{Matrix elements}

The calculation of the matrix elements is much more complex and
requires a detailed expression for each state involved in terms of
operators. We will begin by introducing the notation, using the
ground state as an example. Then we will review the matrix elements
for each process of excitation.

\subsubsection{Notation}

The screened impurity in its ground state is given by the following
Young Tableau represented in figure~\ref{grs}: 
\begin{figure}[here]
\epsfysize=1.5in 
\centerline{\epsfbox{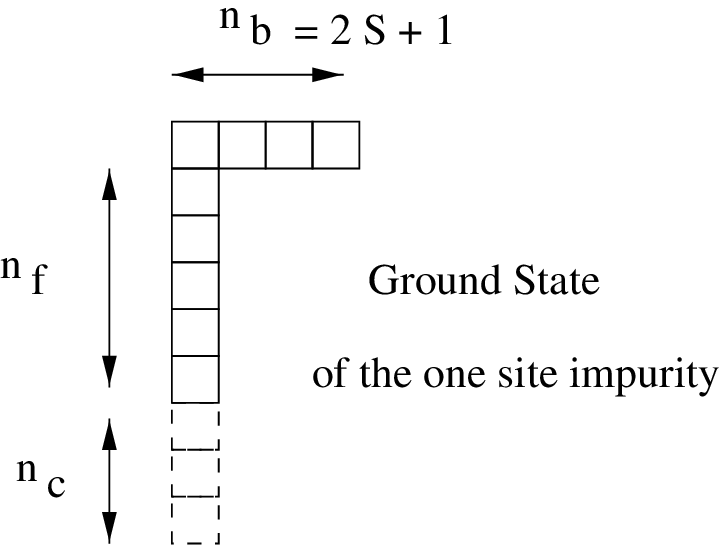}}
\vskip 0.1truein
\protect\caption{ Ground state after the screening of the impurity
spin by conduction electrons. Note that we have changed our notations
here and have given a spin $S + 1/2 $ to the initial impurity spin. }
\label{grs}
\end{figure}
Each box is filled with a field to which an index is attached. In
order to describe the state we have to first symmetrize the indices of the
fields in the row and then antisymmetrize the indices 
in the column. We choose to fill the row with $n_b= 2S$ bosons in a
given
spin state
(say, all the bosons have index A ), and the rest of the column
will be filled with $n_f=n_{f}^{*}-1$ f-fermions and $n_c$ conduction electrons so
that $n_f^{*} + n_c = N$. With these conventions the ground state can be
expressed as
\bea
| GS \rangle  & = & \frac{1}{N_{GS}} \Bigl [ \ \ (b_A^\dagger)^{2S }
\! \! \sum_{  \begin{array}{c} {\scriptstyle {\cal P} \{ \sigma \}  \in } 
 {\scriptstyle \{ 1 \ldots N \} - \{A \} } \end{array}  } \! \!
 \varepsilon ( {\cal P}) \   
 f_{\sigma_1}^\dagger \cdots f_{\sigma_{n_f}}^\dagger
c_{\sigma_{n_f+1}}^\dagger \cdots c_{\sigma_{N-1}}^\dagger | 0 \rangle
 \nonumber \\
 & + &  \! \! \sum_{\sigma_i \neq A} (b_A^\dagger)^{2S-1}
 b_{\sigma_i}^\dagger   \sum_{ 
\begin{array}{c} {\scriptstyle {\cal P}
( \{ \sigma \} ) \in }
{\scriptstyle  ( \{ 1 \ldots N \} - \{ \sigma_i \})}
\end{array} } \! \!   
\varepsilon ( {\cal P}) \  
 f_{\sigma_1}^\dagger \cdots f_{\sigma_{n_f}}^\dagger
c_{\sigma_{n_f+1}}^\dagger \cdots c_{\sigma_{N-1}}^\dagger | 0 \rangle
\Bigr ]
\eea
where $(\sigma_1 \ldots \sigma_n ) = {\cal P} (\alpha_1 \ldots
\alpha_n)$ are permutations of the set of indices $\{ \alpha_1 \cdots
\alpha_n \}$ strictly ordered. $\varepsilon ( {\cal P})$ is the
sign of the permutation. $N_{GS}$ is the normalization factor 
\[
N_{GS} = \sqrt{2S +N -1} \ {N-1 \choose n_f} ^{1/2}  (2S -1 )! \ n_f! \ n_c! \]
where $ {\displaystyle {b \choose a } }$ is the number of ways of chosing $a$
elements out of a group of $b$ possible choices and $n_c= (N-1-n_f)$.
In all that follows we will lighten the notation by keeping track of
the degenerescences in any expression of states. For example the state
$|GS \rangle$ can be written

\bea
\label{nota}
| G S \rangle  & = & \frac{1}{N_{GS}}  \ n_f! \ n_c! \Bigl [ (b_A^\dagger )^{2S-1}
\sum_{  {\scriptstyle {\tilde S}_{n_f}  \in 
 \{ 1 \ldots N \} - \{ A \} }   } \! \! \varepsilon^\prime 
 f_{\sigma_1}^\dagger \cdots f_{\sigma_{n_f}}^\dagger
c_{\sigma_{n_f+1}}^\dagger \cdots c_{\sigma_{N-1}}^\dagger | 0 \rangle
 \nonumber \\
& + &   \sum_{\sigma_i \neq A} (b_A^\dagger)^{2S-1}
 b_{\sigma_i}^\dagger   \sum_{  {\scriptstyle {\tilde S}_{n_f}
 \in  ( \{ 1 \ldots N \} - \{ \sigma_i \})}  } \! \! \varepsilon^\prime    
 f_{\sigma_1}^\dagger \cdots f_{\sigma_{n_f}}^\dagger
c_{\sigma_{n_f+1}}^\dagger \cdots c_{\sigma_{N-1}}^\dagger | 0 \rangle
\Bigr ]
\eea
where we suppose that the two sets of indices $ ( \sigma_1 \ldots
\sigma_{n_f} )$ and $( \sigma_{n_f+1} \ldots \sigma_{N-1} )$
are ordered in a strictly increasing order. $\varepsilon^\prime$ is the sign of the residual permutation of
indices due to the fact that the two sets of indices are not ordered
with respect to each other. $ {\tilde S}_{n_f}  \in \{ 2 \ldots N \}$ denotes the partition of the two sets of indices $ ( \sigma_1 \ldots \sigma_{n_f}) $ and $ ( \sigma_{n_f +1} \ldots \sigma_{N-1} )$ chosen among $ \{2 \ldots N \}$ and strictly ordered inside each set. The important point here is that the above sum runs over linearly
independent states only. Thus when we have to evaluate the norm of a
state, the prefactors in front of the sum have to be squared. We give
in the Appendix the detailed evaluation of the norm of the
ground state as an example.

\subsubsection{ Decomposition rule into parallel and anti-parallel
states}

\begin{figure}
\epsfxsize=0.9\textwidth
\centerline{\epsfbox{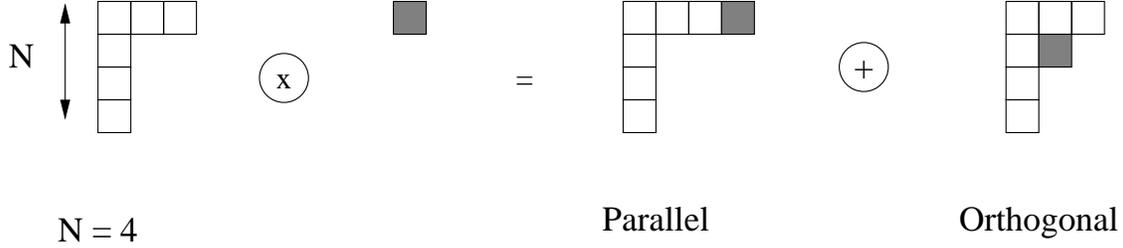}}
\vskip 0.1truein
\protect\caption{Schematic tensor product between the screened
impurity in the ground state and a conduction electron. Note that the
number of boxes in the column of the screened impurity is $N$ (the
impurity is screened).}
\label{tensor}
\end{figure}

When we add an additional electron to the ground-state,
the state that forms is a linear combination of two states,
one in which the  
spin is symmetrically correlated with the impurity, and another in which
or antisymmetrically correlated with the
partially screened impurity, as illustrated by the 
Young tableau ~\ref{tensor}).  In the analoguous SU(2) problem, these
two states would correspond to adding an electron into
a state in which the spin that is either ``parallel'' 
or ``anti-parallel'' to the impurity spin.
Written out in operator language, then if 
$\alpha $ is the spin index of the added electron, 
\bea
\label{decomp}
c^\dagger_\alpha | G S \rangle & = &  \overbrace{ \frac{1}{2S } \left ( c^\dagger_\alpha | GS
\rangle + \left( 2S - 1 \right) c^\dagger_A | \varphi_{int}(\alpha ) \rangle \right ) }^{
\mbox{parallel}} \nonumber \\
    \oplus &  & \overbrace{ \frac{1}{2S } \left ( \left(  2 S -1 \right) \ c^\dagger_\alpha | GS
\rangle - \left( 2S -1\right) c^\dagger_A | \varphi_{int}(\alpha ) \rangle \right )}^{
\mbox{anti-parallel}} \ .
\eea
$|\varphi_{int}(\alpha) \rangle$ is a state where the 
index $\alpha$ of the new box
has been exchanged with the indices of the bosons in the ground state
(here, remember, all the bosons have index $A$ ).
In each case the spin index of the added electron
is symmetrized, or anti-symmetrized with the
the boson spin
indices, leaving the 
indices of the first column 
untouched because this column is a filled singlet in the 
ground state.
The state $|\varphi_{int}(\alpha ) \rangle$ can be written
\bea
|\varphi_{int}( \alpha ) \rangle   =  \frac{1}{N_{GS}} \Bigl [ { N-1 \choose n_f } n_f ! \ n_c! \
(b_A)^{2S -1}  \ b^\dagger_\alpha & &  \sum_{  {\scriptstyle {\tilde S}_{n_f}  \in 
 \{ 1 \ldots N \}- \{A \} }   } \! \! \varepsilon^\prime
 f_{\sigma_1}^\dagger \cdots f_{\sigma_{n_f}}^\dagger
c_{\sigma_{n_f+1}}^\dagger \cdots c_{\sigma_{N-1}}^\dagger | 0 \rangle
 \nonumber \\
  +  \sum_{\alpha_i \neq \{ A, \alpha \} } { N-1 \choose n_f } n_f ! \ n_c! \
(b_A)^{2S -2}  \ b^\dagger_\alpha b^\dagger_{\alpha_i} & & \sum_{  {\scriptstyle {\tilde S}_{n_f}  \in 
 \{ 1 \ldots N \}- \{A \} }   } \! \!  \varepsilon^\prime
 f_{\sigma_1}^\dagger \cdots f_{\sigma_{n_f}}^\dagger
c_{\sigma_{n_f+1}}^\dagger \cdots c_{\sigma_{N-1}}^\dagger | 0 \rangle
 \nonumber \\
 +  { N-1 \choose n_f } n_f ! \ n_c! \
(b_A^\dagger)^{2S-2}  \ (b^\dagger_\alpha)^2  & & \sum_{  {\scriptstyle {\tilde S}_{n_f}  \in \{ 1 \ldots N \} -\{A \} }   } \! \!  \varepsilon^\prime
 f_{\sigma_1}^\dagger \cdots f_{\sigma_{n_f}}^\dagger
c_{\sigma_{n_f+1}}^\dagger \cdots c_{\sigma_{N-1}}^\dagger | 0 \rangle
\ \Bigr ]
\eea
where the first line corresponds to one index $\alpha$ in the bosonic
row of the Young Tableau; the second line corresponds to one indice
$\alpha$ in the bosonic row and one index $\alpha_i \in ( \{1 \ldots N
\} - \{ A, \alpha \} ) $ and the third line to two indices $\alpha$ in the
bosonic row. We note that the state $ | \varphi_{int}(\alpha ) \rangle $ is not
normalized: we have $\langle \varphi_{int}( \alpha ) | \varphi_{int}( \alpha )
\rangle  = 1/ (2 S) $.

It is now convenient to use 
the above decomposition rule~(\ref{decomp}) to define
two initial states $| (GS, 1_A)^{\uparrow\uparrow} \rangle$ and $| (GS, 1_\alpha 
)^{\uparrow \downarrow}\rangle$, where the electron is added in a parallel and antiparallel
spin state, respectively.
$|( GS, 1_A )^{\uparrow\uparrow}\rangle$ is the state where one electron of index $A$ at site 1 is in the parallel configuration with the ground state at site 0. It's thus the projection of $ c^\dagger_{1, A} | GS \rangle $ onto its parallel part. ( Here $c^\dagger_{1,A}$ is the creation operator for an electronat site 1 of index $A$.) We choose to give an index $A$ to the electron at site 1 because it is already in the parallel configuration with the ground state at site 0.
\beq
| (GS, 1_A) ^{\uparrow\uparrow}\rangle \equiv c^\dagger_A | GS \rangle \ .
\eeq
Note that this state is normalized.
$| (GS, 1_\alpha )^{\uparrow \downarrow}\rangle $ is the state where one electron of index $\alpha$ at site 1 is in the anti-parallel configuration with the ground state at site 0. We require $\alpha \neq A$ so that this configuration exists. We want this state to be normalized (by convention) so that
\beq
| (GS, 1_\alpha )^{\uparrow \downarrow}\rangle = \sqrt{\frac{ 2S -1}{2S}} \left [ c^\dagger_{1, \alpha} | GS \rangle - c^\dagger_{1, A} \vert \varphi_{int} (\alpha)\rangle  \right ] \ .
\eeq

\subsubsection{Process 1, parallel case}

In this process, one electron in the initial state $| (GS , 1_A )^{\uparrow\uparrow}\rangle$  makes
a virtual incursion onto site 0 from site 1.  We
get the following expression for the matrix elements:
\beq
\label{eqn 1}
M_{(1)}^{\uparrow\uparrow} = t^2 \sum_{\sigma} \left | \langle GS + 1^{\uparrow\uparrow} , 0 |
c^\dagger_{0, \sigma} c_{1, \sigma} | GS^{\uparrow\uparrow} ,1_A \rangle \right
|^2  \ ,
\eeq
where $c^\dagger_{0, \sigma}$ denotes the creation operator for an electron at site 0 of index $\sigma$ and $c_{1, \sigma}$ denotes the creation operator for an electron at site 1 of index $\sigma$. $\langle GS + 1^{\uparrow\uparrow} , 0 |$ is the intermediate state, with one more electron at site 0 and no electron at site 1. By convention we suppose that both the intermediate and initial states are
normalized. In the numerator in (\ref{eqn 1}), we notice that $
c^\dagger_{0, \sigma} c_{1, \sigma} | GS^{\uparrow\uparrow} ,1_A \rangle = \lambda
| GS + 1^{\uparrow\uparrow} \rangle $, where $\lambda$ is a multiplicative constant. Indeed while the electron is transfered
from site two to site one it stays in the same spin configuration with
respect to the impurity spin. Thus (\ref{eqn 1}) can be rewritten
\beq
M_{(1)}^{\uparrow\uparrow} = t^2  \sum_{\sigma,\sigma^\prime}  \langle 
(GS  ,1_A)^{\uparrow\uparrow} | c^\dagger_{1, \sigma^\prime}  c_{0, \sigma^\prime}
c^\dagger_{0, \sigma } c_{1,\sigma} | (GS,1_A)^{\uparrow\uparrow}  \rangle  \ .
\eeq

As electron at site 1 has index $A$, we get
\[
M_{(1)}^{\uparrow\uparrow} = t^2  \langle GS | c_{0, A} c^\dagger_{0, A} | GS
\rangle \ . \]
Then $ M_{\mathbf 1}^{\uparrow\uparrow} = \langle GS | 1-
c^\dagger_{0, A} c_{0, A} |GS \rangle $ and finally

\beq
\label{eqn 3}
M_{(1)}^{\uparrow\uparrow} = t^2 (1 - \frac{n_c}{2S -1 +N } ) \ ,
\eeq where $ n_c/ (2S -1 + N) $ is the number of conduction electrons of
index 1 in
the ground state. The explicit evaluation can be found in the Appendix.

\subsubsection{ Process 1, anti-parallel case}

The initial state is $| GS^{\uparrow \downarrow} ,1_\alpha \rangle$ where one electron at site 1 is in the anti-parallel configuration with the ground state at site 0.
With the same reasoning as before, we have
\beq
M_{(1)}^{\uparrow \downarrow} = t^2 \sum_{\sigma,\sigma^\prime}   \langle 
(GS,1_{\alpha})^{\uparrow \downarrow}
| c^\dagger_{1,\sigma^\prime} c_{0, \sigma^\prime }
c^\dagger_{0, \sigma } c_{1,\sigma} |(GS,1_{\alpha})^{\uparrow \downarrow} \rangle  \ .
\eeq

Remembering the electron at site 1 has index $\alpha$,
\beq
\label{a2}
M_{(1)}^{\uparrow \downarrow} = t^2  \langle GS | c_{0, \alpha}
c^\dagger_{0, \alpha} | GS \rangle  \ . 
\eeq

Finally, as derived in the Appendix,
\beq
M_{(1)}^{\uparrow \downarrow} = t^2 \left ( 1 - \frac {n_c}{N-1} \right ) \ .
\eeq

\subsection{Process 2, parallel case}

The initial state is $| (GS,1_{\alpha})^{\uparrow \uparrow} \rangle$ where one electron at site 1 with index $A$ is in the parallel configuration with the ground state at site 0. One electron at site 0 jumps onto site 1 and comes back.
\beq
\label{eqn 2}
M_{(2)}^{\uparrow\uparrow} = t^2 \sum_{\sigma} \left | \langle GS - 1^{\uparrow\uparrow} , 2 |
c^\dagger_{1, \sigma} c_{0, \sigma} | (GS, 1_A)^{\uparrow\uparrow} \rangle \right
|^2  \ .
\eeq
As above  we consider that the initial state and the intermediate state $ | GS -1^{\uparrow\uparrow}
, 2 \rangle $ (there is one less electron in the ground state and one more at site 1) are normalized and with the same proportionality
argument,

\bea
M_{(2)}^{\uparrow\uparrow} & = & t^2 \sum_{\sigma, \sigma^\prime} \langle GS ^{\uparrow\uparrow} , 1_A | c^\dagger_{0, \sigma^\prime} c_{1,\sigma^\prime }
c^\dagger_{1, \sigma } c_{0, \sigma} | (GS, 1_A)^{\uparrow\uparrow} \rangle 
 \nonumber
\\
  & = & t^2 \sum_{\sigma,\sigma^\prime} \langle (GS, 1_A)^{\uparrow\uparrow} | c^\dagger_{0,\sigma^\prime}
\left ( \delta_{\sigma,\sigma^\prime} - c^\dagger_{1,\sigma }
c_{1,\sigma^\prime } \right ) c_{0,\sigma} | (GS, 1_A)^{\uparrow\uparrow}\rangle  \ .
\eea
We thus get
\bea
\label{a3}
M_{(2)}^{\uparrow\uparrow} & = & t^2 \sum_{\sigma} \langle 
(GS, 1_A)^{\uparrow\uparrow}| c^\dagger
_{0,\sigma} c_{0,\sigma} |(GS, 1_A)^{\uparrow\uparrow}\rangle \nonumber \\
& - & t^2 \langle GS | c^\dagger_{0, A} c_{0, A} | GS \rangle \ .
\eea
We define
\beq
\label{bla}
M_{(2),\sigma}^{\uparrow\uparrow} \equiv t^2  \langle GS, 1_A | c^\dagger
_{0,\sigma} c_{0,\sigma} |(GS, 1_A)^{\uparrow\uparrow}\rangle  -  t^2 \langle GS | c^\dagger_{0, A} c_{0, A} | GS \rangle \delta_{\sigma, A}
\eeq
so that $ M_{(2)}^{\uparrow\uparrow} = \sum_\sigma M_{(2), \sigma}^{\uparrow\uparrow}$.
The second term in~(\ref{bla}) is the number of conduction electrons in the
ground state that we have already calculated in (\ref{eqn 3}) and
which equals $n_c / ( 2S -1 + N) $. The first term in this equation requires some
exact evaluation with the expression of the state in terms of
operators.
After some algebra done in the Appendix, we get
\beq
M_{(2),\sigma}^{\uparrow\uparrow} = \left \{ \begin{array}{ll}
{\displaystyle t^2 \frac{n_c ( 2S +N -2 ) }{ (N-1) ( 2 S + N-1) }  } \ \ , & \sigma \neq  A
 \\
0 & \sigma = A \end{array} \right . \ .
\eeq
An then
\beq
M_{( 2)}^{\uparrow\uparrow} = t^2 \left ( n_c - \frac{n_c}{2S + N -1} \right ) \ .
\eeq

\subsubsection{Process two, anti-parallel case}

The initial state is $| (GS,1_{\alpha})^{\uparrow \downarrow} \rangle$ where one electron at site 1 with index $\alpha$ is in the anti-parallel configuration with the ground state at site 0. Note that $\alpha \neq A $. On electron at site 0 jumps onto site 1 and comes back.
\bea
M_{(2)}^{\uparrow \downarrow} & = & t^2 \sum_{\sigma,\sigma^\prime} \langle (GS,1_{\alpha})^{\uparrow \downarrow} | c^\dagger_{0, \sigma^\prime} c_{1, \sigma^\prime }
c^\dagger_{1,\sigma } c_{0,\sigma} | (GS,1_{\alpha})^{\uparrow \downarrow} \rangle \nonumber
\\
  & = & t^2  \sum_{\sigma,\sigma^\prime} \langle (GS,1_{\alpha})^{\uparrow \downarrow}| c^\dagger_{0,\sigma^\prime}
\left ( \delta_{\sigma,\sigma^\prime} - c^\dagger_{1,\sigma }
c_{1,\sigma^\prime } \right ) c_{0,\sigma} | (GS,1_{\alpha})^{\uparrow \downarrow} \rangle  \ .
\eea  

Remembering that the electron at site 1 has index $\alpha$ we get
\beq
\label{a4}
M_{(2)}^{\uparrow \downarrow} = t^2  \sum_{\sigma} \langle (GS,1_{\alpha})^{\uparrow \downarrow} |
 c^\dagger_{0, \sigma} c_{0,\sigma} | GS^{\uparrow \downarrow} ,1_\alpha \rangle  - t^2    \langle GS |
 c^\dagger_{0,\alpha }
c_{0,\alpha}  | GS \rangle  \ .
\eeq
We define
\beq
M_{(2),\sigma}^{\uparrow \downarrow} \equiv t^2  \langle (GS,1_{\alpha})^{\uparrow \downarrow}|
 c^\dagger_{0, \sigma} c_{0,\sigma} | (GS,1_{\alpha})^{\uparrow \downarrow}\rangle  - t^2    \langle GS |
 c^\dagger_{0,\alpha }
c_{0,\alpha}  | GS \rangle \delta_{\sigma,\alpha}
\eeq so that $M^{\uparrow \downarrow}_{(2)} = \sum_\sigma M^{\uparrow \downarrow}_{(2),\sigma}$.
After decomposing each state with creation operators, we
get (cf. Appendix) three possibilities depending on the indices.

\beq
M_{(2),\sigma}^{\uparrow \downarrow} = \left \{ \begin{array}{ll}
 {\displaystyle t^2 \frac{n_c}{2 S + N - 1} } & \sigma =A \\
0 & \sigma = \alpha \\
{\displaystyle t^2 \frac{n_c ( 2S +N - 2 ) }{ (N-1) ( 2 S + N - 1) } } \ \ , & \mbox{otherwise}  \end{array} \right . \ ,
\eeq
and after summation upon the indices we find 

\beq
M_{(2)}^{\uparrow \downarrow} = t^2 \left ( n_c - \frac{n_c}{N-1} + \frac{n_c}{(N-1)
(2 S + N - 1) } \right ) \
.
\eeq

\subsubsection{Energy shifts}

We evaluate the energy shifts with the expression

\beq
 \Delta E^{\uparrow\uparrow} -  \Delta E^{\uparrow \downarrow} = \frac{ M^{\uparrow\uparrow}_{(1)} }{
E(n_{e})-E^{\uparrow\uparrow}(n_{e}+1)}  -
 \frac{ M^{\uparrow \downarrow}_{(1)} }{
E(n_{e})-E^{\uparrow \downarrow}(n_{e}+1)} + \frac{ M_{(2)}^{\uparrow \downarrow} -M_{(2)}^{\uparrow\uparrow} }{
E(n_{e})-E(n_{e}-1)}
\eeq
and we get
\bea
\Delta E^{\uparrow\uparrow} -  \Delta E^{\uparrow \downarrow} & = & \frac{t^2}{J} \left [
\frac{n_f N}{ ( N-1 ) ( N -n_f^* -Q/N )} \right . \nonumber \\
& &  - \frac{Q N}{( Q+N 
-n_f^* ) (N +Q +1-2 n_f^* -Q/N )}    \nonumber \\
  &  & \left . - \frac{ ( N -n_f^*) (Q  -n_f^*) N }{ ( n-1) ( Q +N -n_f^* )
  (n_f^* +Q /N) } \right ] \ .
\eea
where we have put $n_f^*=n_f+1$, in keeping with our initial definition
of L-shaped Young tableaux.
This expression is valid for any $ (Q, n_f^*, N) $.

In the large $N$ limit when $ Q/N = q $, $ n_f^*/N = {\tilde n}_f $ and
$(2S+1) /N = {\tilde n}_b $, we get

\beq
\Delta E^{\uparrow\uparrow} -  \Delta E^{\uparrow \downarrow} = \frac{t^2}{J} \left [
\frac{{\tilde n}_f }{1 - {\tilde n}_f } - \frac{ 1- {\tilde n}_f} {
{\tilde n}_f} - \frac{1}{ {\tilde n}_b +1} \left ( \frac{q}{{\tilde
n}_b +1 - {\tilde n }_f } - \frac{1 - {\tilde n}_f} {{\tilde n }_f}
\right  ) \right ] \  .
\eeq

Reducing to the same denominator  and dividing by ${\tilde n}_b $ in  order to
get the effective coupling~(\ref{eq2s}), we finally have
\beq\label{jstar}
J^* = - \frac{t^2}{J (1- {\tilde n}_f ) {\tilde n}_f } \left [ \frac{
\frac{1}{2} - {\tilde n}_f }{ ( 1 - {\tilde n}_f + {\tilde n}_b )} \right ] \ .
\eeq
In the next section, we will check this result in the framework of the
large $ N$ approach, using the formalism of paper~\cite{firstpaper}.

\section{Strong coupling in the large $N$ approach}

  In order to test the formalism developped in~\cite{firstpaper}, we
  need to rederive the effective coupling constant $J^*$ in the large
  $N$ limit with the path integral formulation.

\subsection{Mean-Field theory}

The impurity is described within a $SU(N)$ representation in form of a
L-shaped Young Tableau. The impurity Kondo model is written

\bea
H= \overbrace{\sum_{k, \alpha} \epsilon_{k} c\dg_{k \alpha}c_{k \alpha
}}^{H_o} +\overbrace{\frac{J}{N} c\dg_{\alpha}\pmb{$\Gamma$}_{\alpha \beta}c_{\beta}
\cdot
{\bf S}}^{H_K}
+ \overbrace{
\lambda (Q- Q_0)}^{H_Q}
+ \overbrace{+\zeta ({\cal Y}- Y_o) 
\frac{\zeta}{Q_o}\hat Q \hat { \cal Y}}^{H_{Y}} \ ,
\eea
where $H_o$ describes the conduction electron sea, $H_K$ is
the interaction between the  conduction electron spin density,
 and the local moment, where
$c\dg_{\alpha} = {n_s}^{-1/2} \sum_{k} c\dg_{k\alpha}
$ creates an electron at the site of the local moment ($n_s$ = no. 
of sites). $H_Q$ and $H_Y$
impose the constraints given by $ \hat Q = n_f +n_b = Q_o$ and $\hat {\cal
Y} = n_f - n_b + \frac{1}{Q} \left [ b^\dagger_\alpha f_\alpha = Y_o,
f^\dagger_\beta b_\beta \right ] $. $n_f$ and $n_b$ are respectively the number of
f-electrons and bosons which parametrize the representation.

In the strong coupling limit, this Hamiltonian becomes 
\bea
H & = &  - \frac{J}{N} \sum_{\alpha \beta } f^\dagger_\alpha c_\alpha c^\dagger_\beta f_\beta -
\frac{J}{N} \sum_{\alpha \beta } b^\dagger_\alpha c_\alpha c^\dagger_\beta b_\beta
\nonumber \\
 & + & \lambda (Q- Q_0)+\zeta ({\cal Y}- Y_o) 
.
\eea
After factorizing the Kondo interaction, we obtain
\bea
\label{eqn 6}
H & = & \sum_\sigma \left ( c^\dagger_\sigma \bar V f_\sigma +
f^\dagger_\sigma V c_\sigma \right ) +\sum_\sigma \left ( c^\dagger_\sigma b_\sigma \alpha +
\bar \alpha b^\dagger_\sigma c_\sigma \right ) 
\nonumber \\ 
  & + &  \frac{N}{J} (\bar{V} V- \bar
\alpha \alpha )
\nonumber
 +   \lambda (Q- Q_0)+\zeta ({\cal Y}- Y_o) 
\eea
At the large $N$ saddle point, the fluctuating variable $V$ acquires a
static value, which we take to be $\la V \ra =\la \bar V \ra = V$. The average value of the Grassman
field $\alpha$ is zero, so the bose field is unhybridized in the large
$N$ limit. In order that $n_b\sim 0(N)$, the Bose field must condense
so that the chemical potential of the Bose field, $\lambda_b=\lambda- \zeta$
must vanish, i.e.
$\lambda= \zeta$. The mean-field Hamiltonian is then
\bea
H_{MF} = 
\label{eqn 6b}
H & = & \sum_{\sigma} V\left ( c^\dagger_\sigma  f_\sigma +
f^\dagger_\sigma c_\sigma \right ) + \frac{N}{J} V^2
\nonumber \\ 
  & + &2 \zeta n_f   - 2 \zeta Q_o - \zeta Y_o \ ,
\eea
where the interaction term entering into ${\cal Y}$ is ignored at this
level of approximation. 
After diagonalization of the Fermionic Hamiltonian, we find two
eigenvalues $ E_{\pm} = \zeta \pm \sqrt{\zeta^2 + V^2}$ and the ground
state hamiltonian can be written
\bea
H_{GS}= \sum_{\si } (E_{+}a\dg_{+\ \si }a_{+\ \si } + E_{-}a\dg_{-\ \si}a_{-\  \si}) + 2 N \frac{V^2}{J}- 2 \zeta N \tilde n_f.
\eea
where $a_{\pm \si}= \frac{1}{\sqrt{2}}(c_{\si} \pm f_{\si})$
and ${\tilde n}_f= n_f/N$.  The ground-state corresponds to complete occupation of all $N$ states in the lowest level, 
so that the ground-state energy is 
\beq\label{gse}
E_{GS} = N \left ( \zeta - \sqrt{\zeta^2 +V^2 } + \frac{V^2}{J} - 2 \zeta
{\tilde n}_f \right ) \ .
\eeq
Minimizing $E_{GS}$ with respect to $\zeta $ and $V^2$ gives the two saddle-point equations
\bea
\label{eq50}
1 - \frac{\zeta}{\sqrt{\zeta^2 + V^2}} = 2 {\tilde n}_f \nonumber \\
\frac{1}{2 \sqrt{\zeta^2 +V^2}} = \frac{1}{J} \ ,
\eea 
so that 
\beq
\begin{array}{l}
J= 2 \sqrt{\zeta^2 +V^2} \\
\zeta = J ( 1 - 2 {\tilde n}_f) \end{array} \ .
\eeq
These are the mean-field equations of the strong coupling fixed point.
Substituting back into equation (~\ref{gse}), we obtain
\beq
E_{GS} = - NJ\tilde n_c \tilde n_f
\eeq
where $\tilde n_c= n_c/N = 1- \tilde n_f$ is the number of conduction electrons in the ground-state. 

\subsection{Excitation energies}

 Suppose now that we want to describe the excitation energies
of the large $N$ limit, to compare with those obtained
in a direct strong-coupling expansion~(\ref{shifts}). 
In the mean-field, the energy level $E_-$ is completely filled. 
If we add electrons to the impurity, they must go into the upper
level. We can add a large number of electrons into this level, so we identify
this excitation with the ``anti-parallel'' excitation states, 
giving an excitation energy
\bea
\Delta E^{\perp}= E_+ &=& \zeta + \sqrt{\zeta^2 + V^2}\cr
   &=& J(1- \tilde n_f).
\eea
This matches the large $N$ limit of eq.~(\ref{shifts}).
The leading large N calculation is able to capture the energy to add an
electron into an ``anti-parallel'' configuration because one can 
add many electrons into these states, giving an energy change of order $O(N)$.
By contrast, one can no more than one electron in the parallel
spin configuration, since it is not possible  to symmetrize
more than one electron at a given site.  In this case, the change in energy
is of order $O(1)$, and we must consider the Gaussian corrections to the
mean-field theory to extract this excitation energy.

\subsection{Gaussian fluctuations}

The most important fluctuations about the mean-field theory,
are the fluctuations of the 
$\alpha$-field. The relevant interaction part of the Lagrangian 
is given by 
\bea
  {\cal L}_{\alpha } & = &  \sum_\sigma \left (
 c^\dagger_\sigma b_\sigma \alpha +
\bar \alpha b^\dagger_\sigma c_\sigma \right ) - \frac{N}{J} \bar{\alpha}
\alpha \ .
\nonumber 
\eea
Fluctuations of the alpha field thus mediate an interaction between
the partially screened moment and the conduction electrons, 
given by\\
\vskip 0.05truein
\centerline{\frm{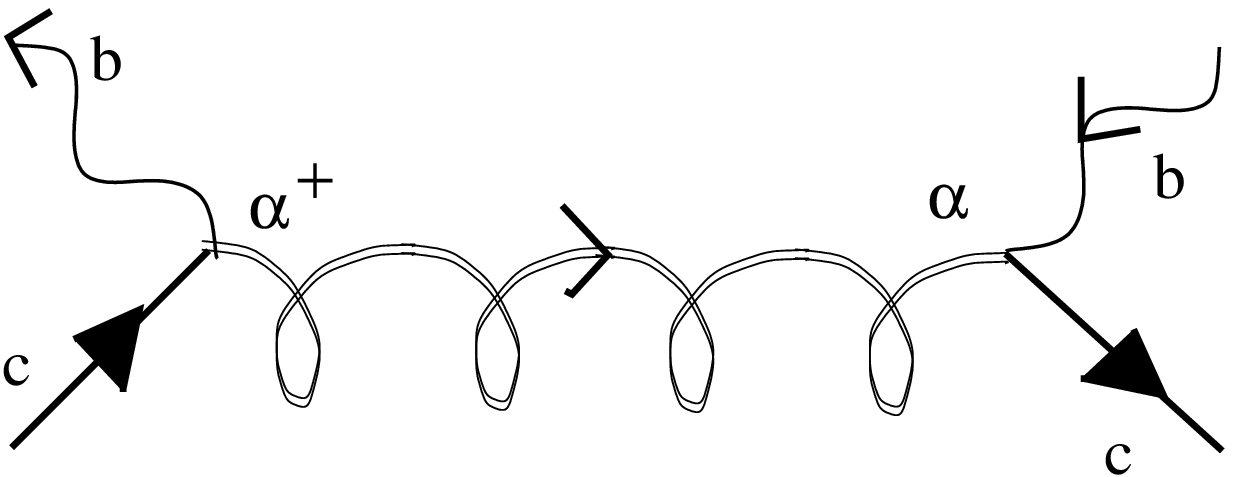}}
\noindent where the full line denotes a conduction electron propagator, 
$G_{c}^{-1} (\omega )= (\omega -\frac{V^{2}}{\omega -2\zeta })$, a wavy line
indicates a boson propagator
$G_{b}^{-1} (\nu )= (\nu - \lambda_{b} )$, where $\lambda _{b}=
\lambda -\zeta $ is the chemical potential of the Bose field and a curly line indicates the
$\alpha $ propagator.  Poles in this propagator describe excitations
associated with the action of the operator $c\dg _{\sigma }b_{\sigma}$
on the ground-state. But since $\langle b_{{\sigma }}\rangle =
\sqrt{n_{b}}\ \delta _{A\sigma }$, this operator describes the addition
of an electron into a parallel spin configuration with the unscreened
local moment.  Thus, to find out the energy to add an electron to the
spin in the ``parallel'' spin configuration, we must look for poles
in the $\alpha $ propagator. 
The action for the $\alpha $ fluctuations is given by
\beq
S_{fluc} = -N \sum_\omega \bar \alpha( \omega ) D^{-1}(\omega)
\alpha(\omega)
\eeq
where
\beq
D^{-1}(\omega) = \frac{1}{J} + T \sum_\nu G_c( \omega + \nu) G_b(\nu)
\ ,
\eeq
Now stationarity of the action with respect to the quantity 
$V$ gives   
\[
\frac{\partial F}{\partial \bar  V }=
\frac{V}{J} + T\sum_{\omega }G_{c} (\omega )\frac{V}{\omega -2\zeta }=0
\]
Using this to replace $1/J$ in the inverse propagator , we obtain
\beq
D^{-1}(\omega) =  T \sum_\nu G_c( \omega + \nu) \left (
\frac{1}{\nu - \lambda_b} - \frac{1}{ \nu +\omega - 2 \zeta} \right ) \
.
\eeq
Carrying out this sum using the contour-integral method, we obtain
\bea
\frac{D^{-1}(\omega)}{(\omega- 2 \zeta)} & = & - \oint \left [ \frac{dz
n_b(z)}{2 \pi i } \frac{1}{ ( z- \lambda_b) ( z+ i \omega_n - E_+ ) (
z+ i \omega_n - E_- )} \right ] \nonumber \\
  & = & - \left [ \frac{n_b(\lambda_b)}{(\omega -E^+ )(\omega- E^-)} +
\sum_{\alpha = \pm} \frac{ - f(E_\alpha)}{  ( E_\alpha -\omega )(
E_\alpha -E_{-\alpha})}  \right ] \nonumber \\
  & = & - \sum_\alpha \frac{{\tilde n}_b -f(E_\alpha)}{(\omega
  -E_\alpha) (E_\alpha -E_{-\alpha}) } \ .
\eea
But at $T=0$, $f(E_-)=1$ and $f(E^+) = 0$ which leads to
\[
D^{-1}(\omega) = \left [ \frac{- {\tilde n}_b }{\omega - E_+} +
\frac{1 + {\tilde n}_b}{ \omega - E_- } \right ] \frac{ ( \omega - 2
\zeta )}{2 \sqrt{V^2 +\zeta^2}} 
\]
so that finally
\beq
D^{-1}(\omega) = \frac{\omega -\omega^*}{(\omega - E_+) (\omega -
E_-)} \frac{(\omega - 2 \zeta )}{2 \sqrt{V^2 +\zeta^2}}
\label{dprop}
\eeq
where $\omega^* = E_+ + J {\tilde n}_b$.
Notice, that $D(\omega)$ has poles at both $\omega  = \omega^*$ and
$\omega = 2 \zeta$.  The appearance of a second pole at $\omega= 2 \zeta$
is a ``ghost'' which factors out of the entire partition function
when we fix the gauge properly\cite{firstpaper}, and it 
does not contribute to physical excitations.
We can identify the excitation energy 
\[
\Delta E_{\perp }=\omega  ^{*}= J ( 1- \tilde{n_{f}}+\tilde{n}_{b})
\]
as the physical energy to add an electron in the
perpendicular spin configuration. This result 
agrees with the large 
$N$ limit of result (22).
result defines the pole in the $\alpha $ propagator.  

\subsection{Renormalized Spin Interaction between  conduction
electrons and residual spin.}

The residual interaction between the partially quenched moment
and the electrons at site ``1'' is given in strong-coupling by 
the diagram
\\
\vskip 0.05truein
\centerline{\frm{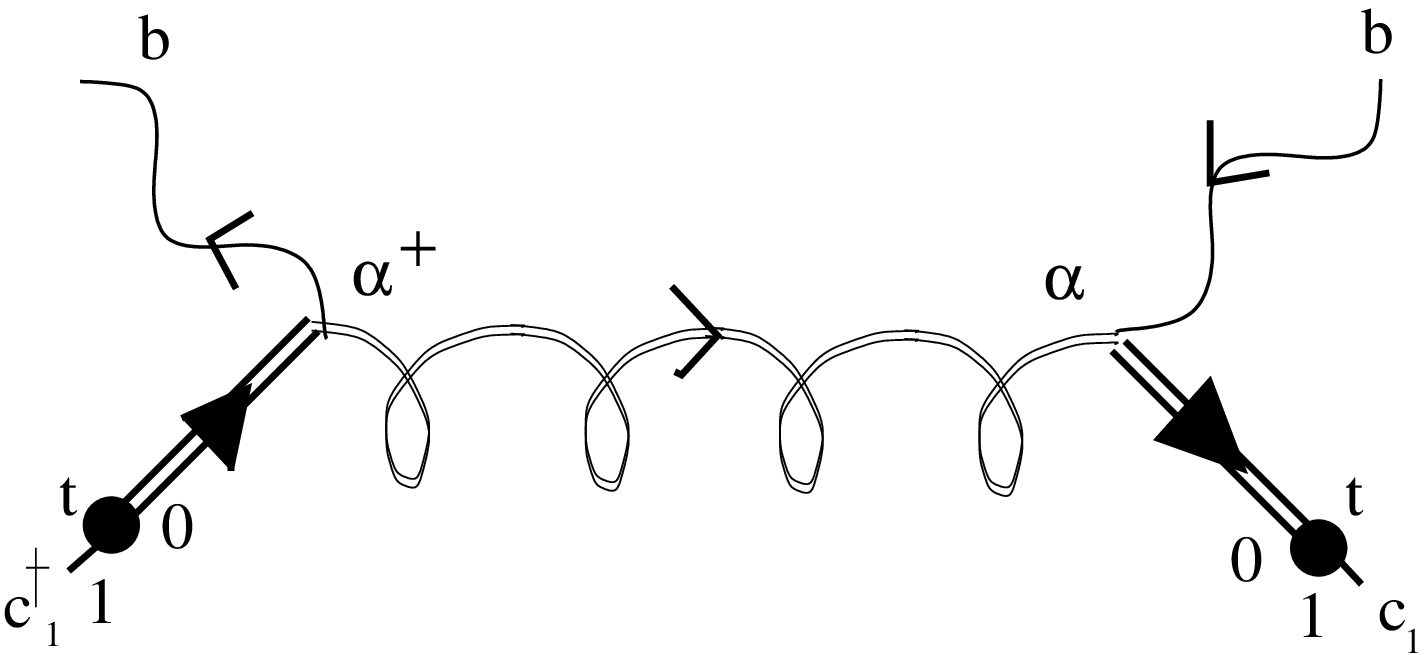}}
\noindent
where the ``$t$'' denotes the hopping matrix element for an
electron moving between site 1 and site 0.  In this diagram,
the external legs do not contribute to the interaction amplitude.
The total interaction strength at low energies is consequenctly 
\beq 
\frac{J^*}{N}= t^2 [G_c(\omega)]^2 D(\omega)\vert_{\omega=0}
\eeq
where
\bea
G_c(\omega)= \frac{1}{\omega - \frac{V^2}{\omega- 2 \zeta}}
\eea
is the propagator for a  conduction electron at site 0. At zero frequency, we have
\bea
G_c(\omega)= \frac{2 \zeta}{V^2}
\eea
From eq (~\ref{dprop}), we have 
\bea
D(0)= -\frac{2 \zeta \omega^*}{J V^2}
\eea
so that
\beq
J^* = - \frac{ Jt^2 (2 \zeta)}{2 \omega^* V^2} \ .
\eeq

Now using the mean-field equations, we get

\[
\begin{array}{l}
\omega^* = J \left (1 - {\tilde n}_f + {\tilde n}_b \right )\\
{\displaystyle \frac{V^2}{2 \zeta} = \frac{1 - 2 {\tilde n}_f}{ J^2 {\tilde n}_f ( 1-
{\tilde n}_f)} }
 \end{array}
\]
so that finally
\beq
J^* = - \frac{t^2 (1/2 - {\tilde n}_f)}{J (1-{\tilde n}_f) {\tilde
n}_f (1-{\tilde n}_f+ {\tilde n}_b )}  \ .
\eeq
This result corresponds to the result obtained using a direct
strong coupling expansion (\ref{jstar}). 

\section{Conclusion}

In this paper we have investigated the effect of the representation in
the description of the SU(N) Kondo model. We used a special class of
Young tableaux which are L-shaped and we obtain a path
integral expression of our model with an additional symmetry between
fermions and bosons: the supersymmetry.
This formalism enables us to tune hte Hund's interactions inside the
representation, which can be used to describe complex atoms like
$U^{3+}$.

We have found that changing the representation gives rise to new fixed
points in the physics of the Kondo model.
When the number of fermions (antisymmetric components in the L-shaped
Young tableau) is larger than $N/2$ (in the large $N$ limit), then a
two-stage Kondo appears, where the impurity spin is screened twice by
two clouds of conduction electrons, leading to a resulting spin of
$S^*=S-1$.
In addition to the two-stage Kondo effect, a class of representations
(where for example $n_b^{*} =2 $ and $n_f^{*} = N-1$ ) lead to a degenerate
ground state and thus to a non fermi liquid. We didn't investigate so
far the properties of this new fixed point.

Alternatively, this paper is a test for the formalism developped in
our previous work about a large $N$ field theory for the SU(N) Kondo
model. We have given here the full renormalization group argument to
prove the unstability of the fixed point leading to the two-stage
Kondo effect. We have matched the unstability criterions in the limit
of large $N$. The excitation energies involved in adding an extra
electron to the screened impurity both in the parallel and anti-parallel
states have been reproduced.

\appendix

\section{Some detailed evaluations}
\subsection{The norm of the ground state}

We illustrate the use of the notation in formula~\ref{nota} by checking
that $ \langle GS | G S \rangle = 1 $.
Using the relation $ \langle 0 | b_A^{2 S} b_A^{\dagger \ 2 S  } |
0 \rangle = (2 S )! $ we have

\bea
\langle GS | G S \rangle & = &\frac{1}{N_{GS}^2} \left [ (2 S )! { N-1 \choose n_f } ( n_f
!)^2  \left ( n_c! \right )^2 \right . \nonumber  \\
 & + & \left . (2 S -1) !  (N-1)  { N-1 \choose n_f } ( n_f
!)^2  \left ( n_c! \right )^2 \right ]
\eea
so that

\beq
\langle GS | G S \rangle = \frac{1}{N_{GS}^2}  ( 2S + N-1 ) (2 S -1 )! { N-1 \choose n_f } ( n_f
!)^2  \left ( n_c! \right )^2  =1
\eeq

\subsection{ Number of conduction electrons of different indices in the ground state}

 \subsubsection{Number of electrons of index 1}
\label{gs}
In the equation~\ref{nota} there are no c-electrons of index 1 in the
first line. We thus get

\beq
c_1 | GS \rangle = \frac{1}{N_{GS}} \sum_{\sigma_i \neq A} (b_A^\dagger)^{2S-1}
 b_{\sigma_i}^\dagger   \sum_{  {\scriptstyle {\tilde S}_{n_f}
 \in  ( \{ 1 \ldots N \} - \{ A,\sigma_i \})}  } \! \! \varepsilon^\prime    
 f_{\sigma_1}^\dagger \cdots f_{\sigma_{n_f}}^\dagger
c_{\sigma_{n_f+1}}^\dagger \cdots c_{\sigma_{N-1}}^\dagger | 0 \rangle
 \ .
\eeq The choice of indices for the f-electrons can not be 1 any more
 because it is attributed to the c-electron in that sum. 
Noting $ n_{c_A}= \langle GS | c^\dagger_A c_A | GS \rangle $ we get

\bea
n_{c_A} & = & {\displaystyle \frac{(N-1) {n_f \choose N-2 } } { (2 S +
N -1 ) { N-1 \choose n_f } }} \nonumber \\
  & = & {\displaystyle \frac{n_c}{(2S +N -1)} } \ ,
\eea where $n_c$ is the total number of c-electrons in the ground
state.

\subsubsection{Number of electrons of index $\sigma \neq A$ in the
ground state}

\bea
c_\sigma \ |GS \rangle  & = & \frac{1}{N_{GS}} \Bigl [ (b_A^\dagger
)^{2S -1} \ n_f! \ n_c!
\sum_{ {\scriptstyle {\tilde S}_{n_f}  \in 
 (\{ 1 \ldots N \}- \{ A, \sigma \} }   } \! \! \varepsilon^\prime 
 f_{\sigma_1}^\dagger \cdots f_{\sigma_{n_f}}^\dagger
  c_{\sigma_{n_f+1}}^\dagger \cdots c_{\sigma_{N-1}}^\dagger | 0 \rangle
 \nonumber \\
& + &   \sum_{\sigma_i \neq \sigma} (b_A^\dagger)^{2S -1}
 b_{\sigma_i}^\dagger   \sum_{  {\scriptstyle {\tilde S}_{n_f}
 \in  ( \{ 1 \ldots N \} - \{ \sigma ,\sigma_i \})}  } \! \! \varepsilon^\prime    
 f_{\sigma_1}^\dagger \cdots f_{\sigma_{n_f}}^\dagger
 c_{\sigma_{n_f+1}}^\dagger \cdots c_{\sigma_{N-1}}^\dagger | 0 \rangle
\Bigr ] \ ; 
\eea

Thus we get

\bea
n_{c \sigma} & = & \frac{1}{N_{GS}^2} \left [ (2 S  + N-2 ) { n_f
\choose N-2} ( 2 S -1 )! \left ( n_f! n_c! \right )^2 \right ]
\nonumber \\ 
  & = & \frac{n_c (2 S + N -2)}{(2S + N - 1) ( N-1)} \ .
\eea

We note that summing over the indices, we can check that the total
number of conduction electrons in the ground state is equal to $n_c= N-n_f-1$.

\subsection{Process 1, anti-parallel case: evaluation of the matrix element}

The electron at site 1 has index $\alpha \neq A$.

\beq
\label{a1}
|GS^{\uparrow \downarrow} , 1_\alpha \rangle  = \sqrt{\frac{2 S -1}{2 S} } \left [
c^\dagger_{1,\alpha} | GS \rangle - c^\dagger_{1, A} | \varphi_{int}(\alpha)
\rangle \right ] \ ,
\eeq
where we have noted $c^\dagger_{1,\alpha} $ the creation operator for an electron of index $\alpha$
at site 1. Due to the presence of the electron at site 1 the two state
in~\ref{a1} are anti-parallel.
We get
\bea
\langle GS^{\uparrow \downarrow}, 1 | GS^{\uparrow \downarrow}, 1 \rangle & = &\frac{2S -1}{2S} \left [ \langle GS | c_{1,\alpha} c_{1,\alpha}^\dagger  |GS \rangle + \langle \varphi_{int}(\alpha) | c_{1,A} c_{1,A}^\dagger | \varphi_{int}(\alpha) \rangle \right ] \nonumber \\
  & = & \frac{2S -1}{2S} \left [  \langle GS | GS \rangle + \langle \varphi_{int}(\alpha) |
\varphi_{int}(\alpha) \rangle \right ] \nonumber \\
 & = & 1 \ .
\eea

Now starting from  eq.~\ref{a2} the full matrix element for this case
is given by

\bea
M^{\uparrow \downarrow}_{(1)} & = & t^2 \frac{2S-1}{2S } \left [ \langle GS | 1- c^\dagger_\alpha
c_\alpha |GS \rangle + \langle \varphi_{int} | 1 - c^\dagger_A c_A |
\varphi_{int} \rangle  + 2 \langle \varphi_{int}| c^\dagger_\alpha c_A
| GS \rangle \right ] \nonumber \\
   & = & t^2 \frac{2 S-1}{2 S } \left [ 1 - \frac { n_c ( 2 S )}{(2 S +
   N-1) (N-1) } + \frac{1 }{2 S-1} \left( 1 - \frac{n_c N}{(2 S + N-1) (N -
   1)} \right ) + 2 \frac{n_c}{(2S+N-1) (N-1) } \right ] \nonumber \\
  & = & t^2 \left ( 1 - \frac{n_c}{N-1} \right ) \ .
\eea

\subsection{Process 2, parallel case: evaluation of the matrix
   element}

We suppose that the electron at site 1 has index A. From eq.~\ref{a3}
it is clear that if the hopping electron had index $\sigma=A$ the
corresponding matrix element vanishes. 
For a hopping electron with index $\sigma \neq A $, we can study the
state $|(GS, 1_A )^{\uparrow\uparrow}\rangle $ in the first term of eq.~\ref{a3}.

We have

\bea
|(GS, 1_A )^{\uparrow\uparrow}\rangle  & = & \frac{1}{N_{GS}} \Bigl [ (b_A^\dagger )^{2S-1} \ n_f! \ n_c!
\sum_{  {\scriptstyle {\tilde S}_{n_f}  \in 
 \{ 1 \ldots N \} - \{A \} }   } \! \! \varepsilon^\prime 
 f_{\sigma_1}^\dagger \cdots f_{\sigma_{n_f}}^\dagger
 \ c^\dagger_{2, A} \ c_{\sigma_{n_f+1}}^\dagger \cdots c_{\sigma_{N-1}}^\dagger | 0 \rangle
 \nonumber \\
& + &   \sum_{\sigma_i \neq A} (b_A^\dagger)^{2S-1}
 b_{\sigma_i}^\dagger   \sum_{  {\scriptstyle {\tilde S}_{n_f}
 \in  ( \{ 1 \ldots N \} - \{ \sigma_i \})}  } \! \! \varepsilon^\prime    
 f_{\sigma_1}^\dagger \cdots f_{\sigma_{n_f}}^\dagger
\ c^\dagger_{2, A}\ c_{\sigma_{n_f+1}}^\dagger \cdots c_{\sigma_{N-1}}^\dagger | 0 \rangle
\Bigr ] \ .
\eea

Thus for $\sigma \neq A$ we get

\bea
c_\sigma \ |(GS, 1_A )^{\uparrow\uparrow}\rangle  & = & \frac{1}{N_{GS}} \Bigl [ (b_A^\dagger )^{2S-1} \ n_f! \ n_c!
\sum_{  {\scriptstyle {\tilde S}_{n_f} \in 
 (\{ 1 \ldots N \}- \{A, \sigma \} }   } \! \! \varepsilon^\prime 
 f_{\sigma_1}^\dagger \cdots f_{\sigma_{n_f}}^\dagger
 \ c^\dagger_{2, A} \ c_{\sigma_{n_f+1}}^\dagger \cdots c_{\sigma_{N-1}}^\dagger | 0 \rangle
 \nonumber \\
& + &   \sum_{\sigma_i \neq A}(b_A^\dagger)^{2S-1}
 b_{\sigma_i}^\dagger   \sum_{  {\scriptstyle {\tilde S}_{n_f}
 \in  ( \{ 1 \ldots N \} - \{ \sigma ,\sigma_i \})}  } \! \! \varepsilon^\prime    
 f_{\sigma_1}^\dagger \cdots f_{\sigma_{n_f}}^\dagger
\ c^\dagger_{2, A}\ c_{\sigma_{n_f+1}}^\dagger \cdots c_{\sigma_{N-1}}^\dagger | 0 \rangle
\Bigr ] 
\eea
and when taking the norm we get

\beq
M^{\uparrow\uparrow}_{ (2), \sigma} = t^2 \frac{n_c ( 2 S + N -2) }{ (N-1) (2S+N-1)} \ \ \ \
{\mbox if} \ \sigma \neq A 
\eeq

\subsection{Process 2, anti-parallel case: evaluation of the matrix element}

In this process we choose on site 1 an electron with an index $\alpha
\neq A $.  The second term in eq.~\ref{a4} corresponds to the number
of c-electrons of index $\alpha \neq 1$ in the ground state which has
been evaluated in section~\ref{gs}. We study the first term in eq.~(\ref{a4}). We have three
different cases.

 \subsubsection{$\sigma \neq \alpha $ and $\sigma \neq A $}

\beq
\label{a5}
c_\sigma |GS^{\uparrow \downarrow},1_\alpha \rangle = - \sqrt{\frac{2 S-1}{2S  }} \left [
c^\dagger_{1, \alpha} c_\sigma |GS \rangle - c^\dagger_{1, A} c_\sigma | \varphi_{int}(\alpha) \rangle \right ] \ .
\eeq

We then have the two states
\bea
c_\sigma \ |GS \rangle  & = & \frac{1}{N_{GS}} \Bigl [ (b_A^\dagger )^{2S-1} \ n_f! \ n_c!
\sum_{  {\scriptstyle {\tilde S}_{n_f}  \in 
 (\{ 1 \ldots N \}- \{ A, \sigma \} }   } \! \! \varepsilon^\prime 
 f_{\sigma_1}^\dagger \cdots f_{\sigma_{n_f}}^\dagger
  c_{\sigma_{n_f+1}}^\dagger \cdots c_{\sigma_{N-1}}^\dagger | 0 \rangle
 \nonumber \\
& + &   \sum_{\sigma_i \neq A} (b_A^\dagger)^{2S-1}
 b_{\sigma_i}^\dagger   \sum_{  {\scriptstyle {\tilde S}_{n_f}
 \in  ( \{ 1 \ldots N \} - \{ \sigma ,\sigma_i \})}  } \! \! \varepsilon^\prime    
 f_{\sigma_1}^\dagger \cdots f_{\sigma_{n_f}}^\dagger
 c_{\sigma_{n_f+1}}^\dagger \cdots c_{\sigma_{N-1}}^\dagger | 0 \rangle
\Bigr ] \ ; 
\eea
and
\bea
c_\sigma |\varphi_{int}(\alpha) \rangle   =  \frac{1}{N_{GS}} \Bigl [ { N-1 \choose n_f } n_f ! \ n_c! \
(b_A)^{2S-1}  \ b^\dagger_\sigma & &  \sum_{  {\scriptstyle {\tilde S}_{n_f}  \in 
 (\{ 1 \ldots N \} - \{A,\sigma \} ) }   } \! \! \varepsilon^\prime
 f_{\sigma_1}^\dagger \cdots f_{\sigma_{n_f}}^\dagger
c_{\sigma_{n_f+1}}^\dagger \cdots c_{\sigma_{N-1}}^\dagger | 0 \rangle
 \nonumber \\
  +  \sum_{\sigma_i \in ( \{ 1 \ldots N \} - \{ A,\sigma \} ) } { N-1 \choose n_f } n_f ! \ n_c! \
(b_A)^{2S -2}  \ b^\dagger_\sigma b^\dagger_{\sigma_i} & & \sum_{  {\scriptstyle {\tilde S}_{n_f}  \in 
 (\{ 1 \ldots N \} - \{ A,\sigma \}) }   } \! \!  \varepsilon^\prime
 f_{\sigma_1}^\dagger \cdots f_{\sigma_{n_f}}^\dagger
c_{\sigma_{n_f+1}}^\dagger \cdots c_{\sigma_{N-1}}^\dagger | 0 \rangle
 \nonumber \\
 +  { N-1 \choose n_f } n_f ! \ n_c! \
(b_A^\dagger)^{2S-2}  \ (b^\dagger_\sigma)^2   \sum_{  {\scriptstyle {\tilde S}_{n_f} \in 
 (\{ 1 \ldots N \} - \{A, \sigma \}) }   } & &\! \!  \varepsilon^\prime
 f_{\sigma_1}^\dagger \cdots f_{\sigma_{n_f}}^\dagger
c_{\sigma_{n_f+1}}^\dagger \cdots c_{\sigma_{N-1}}^\dagger | 0 \rangle
\ \Bigr ] \ .
\eea
For $\sigma \neq \alpha$ these two states are anti-parallel and we get
\bea
\langle GS | c^\dagger_{0,\sigma} c_{1,\alpha} c^\dagger_{1,\alpha} c_{0,\sigma} | GS \rangle  &= & \frac{n_c (2S +N
-2)}{(N-1)(2S+N -1)} \nonumber \\
\langle \varphi_{int}(\alpha) | c^\dagger_{0,\sigma} c_{1,\alpha} c^\dagger_{1,\alpha} c_{0,\sigma} | \varphi_{int}(\alpha) \rangle  &= & \frac{n_c (2S +N
-2)}{2 S (N-1)(2S+N-1)}
\eea
so that we get 
\beq
M^{\uparrow \downarrow}_{ (2), \sigma} = t^2 \frac{n_c (2S +N -2)}{(N-1) (2S+N-1)} \ \ \ \
\mbox{if} \ \sigma \neq \alpha \ \ \mbox{and} \ \sigma \neq A 
\eeq

\subsubsection{$\sigma =A$}

The two states in eq.~\ref{a5} are still anti-parallel in that case.
We have
\beq
c_A \ |GS \rangle  =  \frac{1}{N_{GS}}   \sum_{\sigma_i \neq A} (b_A^\dagger)^{2S-1}
 b_{\sigma_i}^\dagger   \sum_{  {\scriptstyle {\tilde S}_{n_f}
 \in  ( \{ 1 \ldots N \} - \{ A ,\sigma_i \})}  } \! \! \varepsilon^\prime    
 f_{\sigma_1}^\dagger \cdots f_{\sigma_{n_f}}^\dagger
 c_{\sigma_{n_f+1}}^\dagger \cdots c_{\sigma_{N-1}}^\dagger | 0 \rangle
 \ ; 
\eeq

\bea
c_A |\varphi_{int}(\alpha ) \rangle   =  \frac{1}{N_{GS}} \Bigl [
\sum_{\sigma_i \in  \{ 1 \ldots N \}- \{ A \}   } { N-1 \choose n_f } n_f ! \ n_c! \
(b_A)^{2S -2}  \ b^\dagger_\sigma b^\dagger_{\sigma_i} & & \sum_{  {\scriptstyle {\tilde S}_{n_f}  \in 
 \{ 1 \ldots N \} - \{ A \}  }   } \! \!  \varepsilon^\prime
 f_{\sigma_1}^\dagger \cdots f_{\sigma_{n_f}}^\dagger
c_{\sigma_{n_f+1}}^\dagger \cdots c_{\sigma_{N-1}}^\dagger | 0 \rangle
 \nonumber \\
 +  { N-1 \choose n_f } n_f ! \ n_c! \
(b_A^\dagger)^{2S-2}  \ (b^\dagger_\sigma)^2  & & \sum_{  {\scriptstyle {\tilde S}_{n_f} \in 
 \{ 1 \ldots N \}- \{A \}  }   } \! \!  \varepsilon^\prime
 f_{\sigma_1}^\dagger \cdots f_{\sigma_{n_f}}^\dagger
c_{\sigma_{n_f+1}}^\dagger \cdots c_{\sigma_{N-1}}^\dagger | 0 \rangle
\ \Bigr ] \ .
\eea
As a result we get
\beq
M^{\uparrow \downarrow}_{(2), A} = t^2 \frac{n_c}{2S +N -1}
\eeq

\subsubsection{$\sigma = \alpha$}

It is easy from eq.~\ref{a4} to check that in this case $
M^{\uparrow \downarrow}_{\alpha,(2)} =0 $.

\end{document}